\documentclass[aps,prx,twocolumn,footinbib,floatfix,superscriptaddress]{revtex4-2}
\usepackage{lineno}
\usepackage{upgreek}

\usepackage{bm}
\usepackage{graphicx}
\usepackage{indentfirst}
\usepackage{braket}
\usepackage{float}
\usepackage{amsmath}
\usepackage{physics}
\usepackage{CJK}
\usepackage{esint}
\usepackage{color}
\usepackage[T1]{fontenc}
\usepackage{subfigure}
\usepackage{amsfonts}
\usepackage{footmisc}
\usepackage{multirow}
\usepackage[outdir=./]{epstopdf}
\usepackage{svg}

\usepackage{float}
\makeatletter
\let\newfloat\newfloat@ltx
\makeatother
\usepackage{algpseudocode}

\usepackage[english]{babel}
\usepackage{url}
\usepackage{comment}
\usepackage{array}
\usepackage{subfiles}
\usepackage{tikz}
\usetikzlibrary{shapes,arrows}
\usepackage[mathscr]{eucal}
\usepackage[hyperfootnotes=false]{hyperref}
\usepackage{cleveref}
\definecolor{darkblue}{rgb}{0,0,1/2}
\hypersetup{
colorlinks=true,
linkcolor=black,
filecolor=blue,
citecolor=darkblue,  
urlcolor=black,
hyperfootnotes=false
}

\newtheorem{theorem}{Theorem}

\newcommand{\calB}{{\cal B}}

\newcommand{\calE}{{\cal E}}
\newcommand{\bcalE}{{\bm \calE}}

\newcommand{\calS}{{\cal S}}

\newcommand{\calG}{{\cal G}}

\newcommand{\1}{^{(1)}}

\def\revisedfontcolor{black}
\newcommand{\QZ}[1]{{{\textcolor{black}{#1}}}}
\newcommand{\hw}[1]{{{\textcolor{black}{#1}}}}

\def\be{\begin{equation}}
\def\ee{\end{equation}}
\def\ba{\begin{eqnarray}}
\def\ea{\end{eqnarray}}
\def\bal{\begin{equation}\begin{aligned}}
\def\eal{\end{aligned}\end{equation}}

\usepackage[normalem]{ulem}
\begin{document}

%TC:ignore

\title{{Overcoming the fundamental limit of \\
quantum transduction via intraband entanglement}}
\begin{abstract}
A quantum transducer converts an input signal to an output probe at a {distant frequency band} while maintaining the quantum information with high fidelity, {which is crucial for quantum networking and distributed quantum sensing and computing.} 
{In terms of microwave-optical quantum transduction}, the state-of-the-art quantum transducers suffer low transduction efficiency from weak nonlinear coupling, wherein increasing pump power to enhance efficiency inevitably leads to thermal noise from heating.
Moreover, we reveal that the efficiency-bandwidth product of a {cavity {electro-optical or electro-optomechanical} transducer} is fundamentally limited
by pump power and nonlinear coupling coefficient, 
irrespective of cavity engineering efforts.
To {overcome this fundamental limit}, we propose to {noiselessly} boost the transduction efficiency by consuming {intraband} entanglement (e.g., microwave-microwave or optical-optical entanglement in the case of microwave-optical transduction). Via a squeezer-coupler-antisqueezer sandwich structure, the protocol enhances the transduction efficiency to unity in the ideal lossless case, given an arbitrarily weak pump and nonlinear coupling. In practical cavity systems, our entanglement-assisted protocol surpasses the non-assisted fundamental limit of the efficiency-bandwidth product and reduces the threshold cooperativity for positive quantum capacity by a factor proportional to two-mode squeezing gain. Given a fixed cooperativity, our approach increases the broadband quantum capacity by orders of magnitude. \QZ{The entanglement-assisted advantage is robust to ancilla loss and cavity detuning.}
\end{abstract}

\author{Haowei Shi}
\email{hwshi@usc.edu}
\address{
Ming Hsieh Department of Electrical and Computer Engineering, University of Southern California, Los
Angeles, California 90089, USA
}

\author{Quntao Zhuang}
\email{qzhuang@usc.edu}

\address{
Ming Hsieh Department of Electrical and Computer Engineering, University of Southern California, Los
Angeles, California 90089, USA
}
\address{
Department of Physics and Astronomy, University of Southern California, Los
Angeles, California 90089, USA
}

\maketitle
%TC:endignore

\section{Introduction}
Quantum transduction aims to interconnect quantum computers and processors via converting quantum states between different frequencies~\cite{lauk2020perspectives,awschalom2021development,han2021microwave}. It serves as the hinge between the microwave superconducting qubits and the optical telecommunication photons, enabling robust quantum networking~\cite{Acin2007,kimble2008quantum,wehner2018quantum,kozlowski2019towards}, and ultimately distributed quantum sensing~\cite{zhang2021distributed} and distributed quantum computing~\cite{monroe2014large,barz2012demonstration}.
Despite the proposals based on various physical platforms~\cite{andrews2014bidirectional,bochmann2013nanomechanical,vainsencher2016bi,balram2016coherent,Tsang2010,Tsang2011,fan2018superconducting,xu2020bidirectional,jiang2020efficient,PhysRevLett.103.043603,PhysRevLett.113.203601,shao2019microwave,fiaschi2021optomechanical,han2020cavity,zhong2020proposal,mirhosseini2020superconducting,forsch2020microwave}, current quantum transduction systems are still far from satisfying, hurdled by a conundrum to balance transduction efficiency, pump-induced heating, and bandwidth~\cite{holzgrafe2020cavity,mirhosseini2020superconducting,sahu2022quantum,brubaker2022optomechanical,qiu2023coherent,sahu2023entangling}. 

An ideal transducer has unity transduction efficiency, zero added noise, and large bandwidth. {As one-way quantum communication is forbidden for efficiency below $50\%$~\cite{wolf2007quantum}, remarkable efforts have been made to improve the on-resonance transduction efficiency to $>50\%$. For example, the recent progress in electro-optomechanical transducers~\cite{andrews2014bidirectional,higginbotham2018harnessing} adopts extremely high-Q mechanical resonators as a mediating mode to connect the microwave mode and the optical mode, which achieves the highest transduction efficiency up to $47\%$ so far with 3.2 noise photons~\cite{brubaker2022optomechanical}. However, such mediation boosts the on-resonance efficiency at the cost of bandwidth, e.g. the bandwidth is limited to $2$ kHz in Ref.~\cite{brubaker2022optomechanical}, c.f. typical bandwidth $\sim 10$MHz of direct conversion~\cite{hease2020bidirectional,holzgrafe2020cavity,sahu2022quantum}. 
Indeed, the bandwidth of an electro-optomechanical transducer is limited below the mechanical resonance frequency $\sim$MHz, which must operate at the resolved sideband limit to suppress the undesired blue sideband two-mode squeezing noise~\cite{brubaker2022optomechanical,caves1982quantum}.
GHz piezo-optomechanical transducers~\cite{jiang2020efficient,mirhosseini2020superconducting} offer room-temperature broadband transduction, but the transduction efficiency is limited, e.g. $\sim 10^{-5}$ in Refs.~\cite{jiang2020efficient,mirhosseini2020superconducting}, due to the optical absorption heating of mechanical resonators~\cite{meenehan2014silicon}. The direct electro-optical conversion~\cite{Tsang2010,Tsang2011} is free from the complications due to the mechanical mode, whereas its transduction efficiency is still limited~\cite{fan2018superconducting,holzgrafe2020cavity}. 
Pulsed pumping has been demonstrated to mitigate the heating and further boost the instantaneous nonlinear coupling for piezo-optomechanical transduction \cite{mirhosseini2020superconducting} and direct electro-optical transduction~\cite{hease2020bidirectional,sahu2022quantum}, however its low duty cycle drastically reduces the transduction rate and it is incompatible with continuous-wave signals. }  

{Such a tradeoff between the transduction efficiency and bandwidth is inevitable. In this paper, we reveal that the efficiency-bandwidth product~\cite{zhao2024quantum} (EBP) of cavity electro-optical or electro-optomechanical transduction, and any transduction with similar Hamiltonian,} is fundamentally limited by the nonlinear coupling coefficient and pump amplitude, regardless of the linewidths of cavities. 
Unfortunately, the nonlinear coupling between photons is intrinsically weak, and a stronger pump inevitably induces more thermal noises~\cite{meenehan2014silicon}.
Therefore, besides the endeavor in materials science and nanofabrication, paradigm shifts are needed to {boost quantum transduction and overcome the limit}.

{Recently, there have been theoretical efforts towards this goal. For example, one can utilize the conventionally discarded environment output to correct the transducer imperfection, via adaptive control~\cite{zhang2018quantum} or Gottesman-Kitaev-Preskill (GKP) encoding~\cite{wang2024passive}. However, the adaptive control protocol relies on ultra-precise broadband homodyne measurement and adaptive displacement in addition to inline squeezing;
the state-of-the-art systems for GKP state engineering~\cite{konno2024logical} are far from usable. In addition, the GKP qubit encoding of the input quantum information is not compatible with continuous-variables. Other approaches rely on crossband microwave-optical entanglement~\cite{sahu2023entangling,meesala2024non} to enable the teleportation-based transduction approach~\cite{wu2021deterministic,zhong2020proposal}; however, noiseless teleportation requires high fidelity crossband entanglement and thus extremely high pump power along with the heating issue as challenging as the direct frequency conversion.}

\begin{figure}[t]
    \centering
    \includegraphics[width=\linewidth]{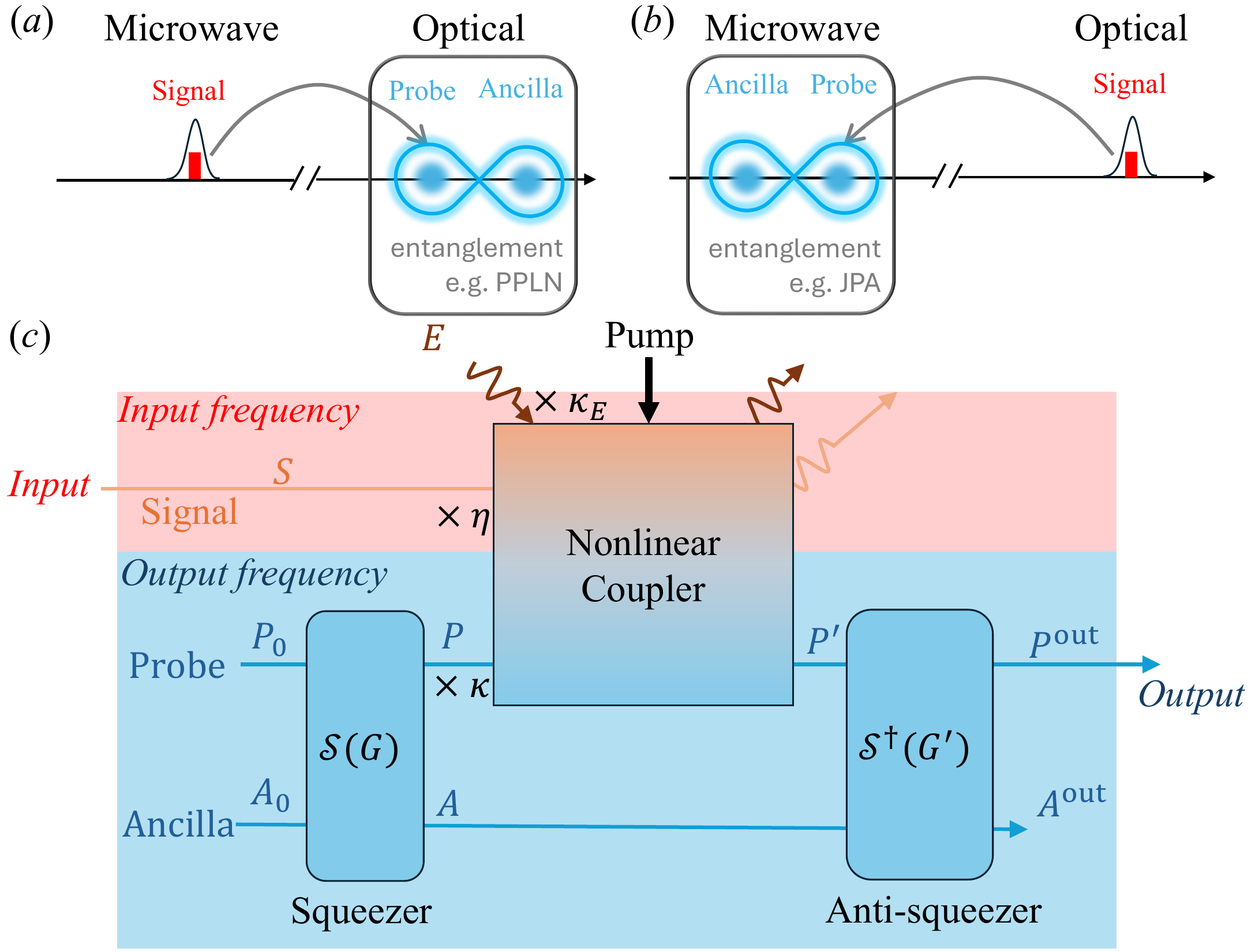}
    \caption{Schematic of the entanglement-assisted (EA) transduction protocol. {(a) EA microwave-to-optical transduction, enhanced by optical entanglement which can be generated by periodically-poled lithium niobate (PPLN)~\cite{nehra2022few}. (b) EA optical-to-microwave transduction, enhanced by microwave entanglement which can be generated by Josephsone parametric amplifier (JPA)~\cite{backes2021quantum,xu2023magnetic,qiu2023broadband}.}
    (c) Detailed protocol. An input signal $S$ is converted to an output probe $P^{\rm out}$ at a different frequency.
    {The probe and ancilla are initially cooled to vacuum state, and two-mode squeezed by $\calS(G)$ of gain $G$. Then, the signal is cast onto the probe by a nonlinear coupler, which is modelled as a beamsplitter of transmissivity $\kappa$, signal-probe conversion efficiency $\eta$, and loss $\kappa_E=1-\kappa-\eta$. Finally the probe and ancilla are antisqueezed by $\calS^\dagger(G')$, the probe is output while the ancilla is discarded. } 
    %The protocol is boosted by the entanglement at the output frequency band between the probe $P$ and the ancilla $A$, which is generated by a two-mode squeezer $\calS(G)$ of gain $G$ prior to entering the nonlinear coupler. Then the input signal is converted to the probe frequency via a nonlinear coupler, while an environment $E$ is inevitably mixed in. After the nonlinear coupling, an antisqueezer $\calS^\dagger(G')$ of gain $G'$ amplifies the signal, while the noise background contributed by $P_0$ and $A_0$ is nulled to the ground state (vacuum).
    }
    \label{fig:schematic}
\end{figure}

{ 
In this work, we propose an intraband-entanglement-assisted protocol to achieve a noiseless broadband enhancement in the efficiency of bosonic transduction between arbitrary distant frequencies, therefore overcome the fundamental limitation on EBP. Adopting techniques from entanglement-assisted (EA) weak signal sensing~\cite{shi2023ultimate} and nonlinear interferometry~\cite{ou1993quantum,chekhova2016nonlinear}, the proposed protocol only requires intraband (optical-optical or microwave-microwave) entanglement as shown in Fig.~\ref{fig:schematic}(a)(b), distinct from teleportation which requires crossband entanglement, measurement and conditional operation.}
In the absence of loss, for an arbitrarily weak nonlinear coupling, the transduction efficiency can always be enhanced up to unity without any added noise. In the next section, we provide an overview of the protocol and its EA advantage.
% {While the mechanism of entanglement assistance applies to general bosonic transduction, in this paper we focus on the direct electro-optical conversion for its simplicity.}
% With broadband cavity electro-optic model taken into account, we show that our protocol goes beyond the EBP limit of traditional cavity-enhanced transducers by a factor increasing with the two-mode squeezing gain $G$ of the intraband entanglement. For quantum communication, the broadband one-way quantum capacity is increased by orders of magnitude, lowering the threshold cooperativity for positive quantum capacity by a factor of $\sim 1/G$. 

{\section{Overview}}

{Entanglement assistance is known as a powerful resource that enhances the precision of weak signal detection beyond the standard quantum limit (SQL) in various scenarios, e.g. nonlinear interferometry~\cite{hudelist2014quantum,chekhova2016nonlinear}, quantum illumination radar~\cite{Guha2009} and dark matter search~\cite{wurtz2021cavity,shi2023ultimate}, via combining two-mode squeezing and antisqueezing before and after the sensing process. The EA advantage has been demonstrated experimentally using photonic ancilla~\cite{hao2022demonstration,jiang2023accelerated} and spin ancilla~\cite{gilmore2021quantum}. In this paper, we exploit entanglement assistance to boost quantum transduction.
}

{
The EA protocol is shown in Fig.~\ref{fig:schematic}. Our protocol features an ancilla entangled with the output at the same frequency band, e.g. at optical or microwave band for microwave-optical transduction as shown in subplots (a)(b). The entanglement is generated via intraband two-mode squeezer $\calS$ before the traditional nonlinear coupling, and processed by an antisqueezer ${\cal S}^\dagger$ afterwards, as shown in subplot (c).
Such intraband entanglement is much easier to implement than crossband entanglement required in teleportation-based transducers~\cite{wu2021deterministic,zhong2020proposal}. 
For example, microwave squeezers have been well established via Josephson parametric amplifier (JPA)~\cite{backes2021quantum,xu2023magnetic,qiu2023broadband}. Optical entanglement has been readily generated using potassium titanyl phosphate~\cite{eberle2013stable} and periodically-poled lithium niobate (PPLN)~\cite{nehra2022few}, while optical inline squeezers are also being actively developed~\cite{yan2012cascaded,wang2022experimental}. While we explicitly consider optical-microwave transduction in subplots (a)(b), in subplot (c) we choose not to specify the input and output frequencies in our protocol, since the protocol allows general bosonic transduction, including phonon-photon conversion~\cite{safavi2011proposal}.
}

{
The pumped nonlinear coupler can be described by a linearized input-output relation, specifically a two-mode bosonic Gaussian channel which can be categorized into beamsplitter-type or two-mode-squeezing-type depending on the pump detuning, for both the cavity electro-optic coupling~\cite{Tsang2011} and the cavity electro-optomechanical coupling~\cite{rau2022entanglement}. Our analysis focuses on beamsplitter-type nonlinear couplers, of which the pumps are red-detuned, which avoid the two-mode squeezing noise~\cite{brubaker2022optomechanical,caves1982quantum} and allow noiseless quantum transduction.
}

%TC:ignore
%TC:endignore

The performance of noiseless quantum transduction is characterized by the signal-to-probe photon conversion efficiency $\eta(\omega)$, as a function of the signal frequency $\omega$. The broadband performance can be quantified by the EBP 
\be 
\calB \equiv \int_{-\infty}^\infty\eta(\omega) d\omega
\label{def_EBP}
\ee 
or the broadband quantum capacity~\cite{lloyd1997capacity,shor2002quantum,devetak2005private,holevo2001evaluating,wang2022quantum} 
\be 
Q_1=\int \frac{d\omega}{2\pi} \max\left[\log_2\left(\frac{\eta\left(\omega\right)}{1-\eta\left(\omega\right)}\right),0\right],
\label{def_Q}
\ee 
of which fundamental limits can be proven for beamsplitter-type quantum transducers. In terms of EBP, we prove fundamental limits in Section~\ref{sec:fundamental_EBP} (see Theorem~\ref{theorem:EO} and Theorem~\ref{theorem:EMO}), as summarized in Theorem~\ref{theorem:summary}.
\begin{theorem}
(informal overview) The EBP of an electro-optical transducer $\calB$ is upper bounded by $\calB \le \pi |g \alpha|$, limited by the nonlinear coupling coefficient $g$ and in-cavity pump power $|\alpha|^2$, regardless of cavity linewidths. 
Enhanced by a mechanical mediating mode, the EBP of an electro-optomechanical transducer is still upper bounded similarly by nonlinear coupling coefficients and pump power. 
\label{theorem:summary}
\end{theorem}
Such fundamental limits hold for any nonlinear couplings of similar Hamiltonians.
In terms of quantum capacity, it is known that cooperativity of electro-optical transduction cavity needs to overcome a threshold $C_{\rm th}=3-2\sqrt{2}$ to enable any non-zero capacity~\cite{wu2021deterministic}. 
These fundamental limits create a conundrum in balancing pump power and heating in quantum transduction engineering.

Our main result is that these fundamental limits can be overcome by utilizing intraband entanglement. In the simple beamsplitter model of nonlinear coupling, the squeezer-coupler-antisqueezer protocol allows noiseless amplification~\cite{ou1993quantum} of the nonlinear coupling, capable of boosting an arbitrarily low transduction efficiency to unity. Such entanglement-assisted noise reduction in parametric amplification has been demonstrated in experiments~\cite{kong2013cancellation,guo2016quantum}. In the full cavity model, in terms of the quantum capacity, the cooperativity threshold can be lowered by a factor of $C_{\rm th, EA}\sim 1/G$ proportional to the two-mode squeezing gain $G$, relaxing the requirement of cavity engineering drastically. In terms of EBP, the proposed entanglement-assisted transduction protocol enables $\calB_{\rm EA}\sim G\cdot \calB\,$, allowing a factor of $G$ advantage in EBP.

Our paper is organized as the following.
We begin with the simple beamsplitter model of coupling in Section~\ref{sec:simplified} to introduce the core mechanism of the protocol and analyze the advantage at a single frequency. Then, we connect the beamsplitter model to the physical cavity model in Section~\ref{sec:full_cavity_model}, where we derive the fundamental limits on transduction and show these limits can be overcome by intraband entanglement. \hw{Several appendices addressing robustness of our protocol to experimental imperfections are noteworthy: Appendix~\ref{app:ancilla_loss} addresses losses in two-mode squeezing operations which can be simplified to ancilla storage loss, where robustness to loss is observed; Appendix~\ref{app:full_cavity_model_EO} addresses imperfect pump detunings, where robustness to large detuning is identified; Appendix~\ref{app:freq_dependent_antisqz} addresses the implementation of frequency-dependent squeezing, where a sequential array of cavity parametric amplifiers is shown to approach the required squeezing spectrum.}

\section{Beamsplitter model of coupling}
\label{sec:simplified}

\subsection{Protocol design}
\label{sec:protocol}
As shown in Fig.~\ref{fig:schematic}(c), a general bosonic transducer converts an input signal $S$ to an output probe $P'$ at different frequency bands via a nonlinear coupler. A general model for \hw{such coupler} is a frequency-dependent beamsplitter~\cite{Tsang2011,rau2022entanglement,andrews2014bidirectional,higginbotham2018harnessing}. {Given a specific input frequency, the coupling can be modelled by a beamsplitter~\cite{Weedbrook2012}. Without entanglement assistance, the transduction efficiency is limited by the signal-probe photon conversion efficiency $\QZ{0<}\eta\hw{\ll 1}$.} The transmissivity of the initial probe $P$ is $\kappa\le 1-\eta$, as an environment port $E$ is inevitably mixed in with transmissivity (the intrinsic loss) $\kappa_E=1-\kappa-\eta\ge 0$. 
Its input-output relation in Fourier frequency domain is
\bal 
\hat \calE_{P'}=e^{i\theta_P}\sqrt{\kappa} \hat \calE_P+e^{i\theta_S}\sqrt{\eta} \hat \calE_{S}+\sqrt{\kappa_E}\hat \calE_E,
\label{eq:channel}
\eal 
where $\theta_P$, $\theta_S$ are phase shifts during the coupling. Here $\hat{\calE}_X(\omega)$ is the traveling-wave field operator of system $X$ at frequency $\omega$ relative to its own carrier, satisfying the commutation relation $[\hat{\calE}_X(\omega),\hat{\calE}_X^\dagger(\omega')]=\delta(\omega-\omega')$. In this section, we focus on the beamsplitter model at a single frequency and omit $\omega$ for simplicity, while the broadband case will be discussed later in the full cavity model.
We will connect such a {beamsplitter model} to {the physical cavity electro-optics and electro-optomechanical systems} in Section~\ref{sec:full_cavity_model}.

{To enhance the overall quantum transduction efficiency from the input to the output, we} amplify the signal-carrying probe while keeping the noise background in vacuum state. To suppress the noise, we introduce an ancilla $A$. The ancilla and the probe runs a `squeezer-coupler-antisqueezer' protocol with a sandwich structure for the transducer: first the probe $P$ and the ancilla $A$ are {cooled to vacuum states} and entangled by a two-mode squeezer with gain $G$; then, a portion $\eta$ of the signal $S$ is converted to the probe $P'$ via nonlinear coupling; finally, the converted probe $P'$ and the ancilla $A$ are antisqueezed with gain $G'$ to produce the final converted output $P^{\rm out}$. The squeezer and the antisqueezer are set to null the probe back to vacuum when the input is vacuum. \hw{In the main text, we ignore the loss in the squeezer and antisqueezer, as they operate on the probe-ancilla pair at close frequencies (e.g. both in the microwave frequencies~\cite{backes2021quantum,xu2023magnetic,qiu2023broadband}) thus much easier to engineer than the signal-probe coupler. We analyze the impact of squeezing loss in Appendix~\ref{app:ancilla_loss}, which shows that the advantage of our EA protocol is robust against the loss in squeezing and ancilla storage.}

Below, we elaborate this protocol step by step.
Before the signal-probe coupling, we prepare the probe and the ancilla using a two-mode squeezer $\calS(G)$ of gain $G$ on initial vacuums $P_0$ and $A_0$. The input-output relation can be conveniently expressed via the linear transform of the field operators
\bal 
\hat \calE_P=\sqrt{G} \hat \calE_{P_0}+\sqrt{G-1} \hat \calE_{A_0}^{\dagger }\,,\\
\hat \calE_A=\sqrt{G-1} \hat \calE_{P_0}^{\dagger }+\sqrt{G} \hat \calE_{A_0}\,.
\eal 
After the two-mode squeezing, the signal is coupled to the probe via the nonlinear coupler by \eqref{eq:channel}. 
Finally, the probe and the ancilla are antisqueezed using $\calS^\dagger (G')$ to output
\bal 
\hat \calE_{P^{\rm out}}=e^{-i\theta_P}\sqrt{G'} \hat \calE_{P'}-\sqrt{G'-1} \hat \calE_{A}^{\dagger }\,,
% \hat \calE_{A^{\rm out}}=-e^{i\theta_P}\sqrt{G'-1} \hat \calE_{P^\prime}^{\dagger }+\sqrt{G'} \hat \calE_{A},
\label{eq:output}
\eal 
where the phase is chosen to cancel the transduction phase shifts $\theta_P$ in \eqref{eq:channel}. {The full formula of the overall input-output relation can be found as \eqref{eq:outputprobe_general_app} in Appendix~\ref{app:input-output}.
To minimize the transduction noise, we solve $G'$ to keep the output to vacuum when the input signal is vacuum, which gives
\be 
G'\gets G^{\prime\star}\equiv \frac{1}{1-\kappa+\kappa/G}.
\label{eq:optGp_nonideal}
\ee
In this case, the noise background in the output probe is vacuum
\be 
\hat \calE_{P^{\rm out}}^\star=
\sqrt{\eta_{\rm EA}} e^{i(\theta_S-\theta_P)}\hat \calE_S + \sqrt{1-\eta_{\rm EA}}\hat \calE_{\rm VAC} \,,
\label{eq:input-output}
\ee
where the background $\hat \calE_{\rm VAC}$ is in vacuum state, and
\be 
\eta_{\rm EA}= \eta G^{\prime\star} =\eta \frac{G}{G (1-\kappa )+\kappa },
\label{eq:etaEA_full}
\ee
is the noiseless EA transduction efficiency. It is noteworthy that the transduction efficiency enhancement holds even if the probe and ancilla are initially in thermal states, wherein the background $\hat \calE_{\rm VAC}$ will be in a thermal state instead. }

One can regard the two-mode antisqueezer $\calS^\dagger(G')$ as an amplifier of the probe, while the first input two-mode squeezer $\calS(G)$ reduces the amplifier noise~\cite{ou1993quantum,kong2013cancellation} from the antisqueezer. \eqref{eq:optGp_nonideal} indicates that to increase the signal amplification $G'$ noiselessly, the input squeezer gain $G$ needs to increase accordingly to suppress the amplification noise. Below, our analysis begins with the ideal lossless case where $\kappa_E=0$ and then proceeds to the lossy case of $\kappa_E>0$.

\subsection{ Lossless coupler: unity-efficiency transduction}
Now we assume the lossless limit $\kappa_E= 0$ to gain intuition about the protocol design, which is {always true} at the cavity overcoupling limit (see Section~\ref{sec:full_cavity_model}). In this case, $\kappa=1-\eta$ and the optimal gain in \eqref{eq:optGp_nonideal} reduces to 
$
G'\gets G^{\prime\star}\equiv {1}/{[\eta+(1-\eta)/G]}\,.
% \label{eq:optGp_ideal}
$
The EA transduction efficiency is
\be 
\eta_{\rm EA}|_{\kappa_E=0}= \eta \cdot \frac{G}{G \eta+(1-\eta) }\,,
\label{eq:etaEA_ideal}
\ee
which approaches unity in the strong squeezing limit,
\be 
 \eta_{\rm EA}|_{\kappa_E=0}\to 1, \mbox{when $G^\prime\to 1/\eta$ and $G\to\infty$}.
\label{eq:uniteff}
\ee
At this limit, the output probe
$\hat \calE_{P^{\rm out}}^\star= e^{i(\theta_S-\theta_P)}\hat \calE_S$ is reflectionless in both quadratures. \hw{Although the reflectionless transduction requires infinite squeezing in \eqref{eq:uniteff}, the EA advantage is still significant at finite squeezing.} For a finite gain, the EA protocol increases the efficiency to $\eta_{\rm EA}\simeq G\eta$ by the amplifier gain factor $G$, at the weak nonlinear coupling limit $\eta\ll1$. Note that here no-cloning~\cite{wootters1982single} is not violated because the other output of the nonlinear coupler is infinitely noisy at the $G\to\infty$ limit. \hw{The intuition behind such an enhancement is that the ancilla $A$ provides a reference entangled with the quantum fluctuation in the probe $P$ after the two-mode squeezer $\calS(G)$. At the limit $G\to \infty$, the quadratures of the two modes are fully correlated as $\Re \hat \calE_P=\Re \hat \calE_A,\Im \hat \calE_P=-\Im \hat \calE_A$~\cite{Weedbrook2012}. Thus the antisqueezer $\calS^\dagger (G')$ can noiselessly amplify the signal transduced into $P^\prime$, where the quantum noise from $A$ during the antisqueezing interference can be completely cancelled utilizing the remaining entanglement. At the lossless limit, the signal is perfectly recovered.}

To enable quantum communication with one-way quantum capacity $Q_1>0$, one needs the overall conversion efficiency above the \emph{zero-quantum-capacity threshold}, $\eta_{\rm EA}>1/2$~\cite{wolf2007quantum}, leading to $\eta>1/(G+1)$ which is drastically easier to achieve than the non-EA case of $\eta>1/2$.

\subsection{ Lossy coupler}
Here we consider general case with intrinsic loss $\kappa_E>0$. In the strong squeezing limit, the EA transduction efficiency \eqref{eq:etaEA_full} goes to 
\be 
\eta_{\rm EA}\to  \frac{1}{1+\kappa_E/\eta}, \mbox{when $G^\prime\to 1/(1-\kappa)$ and $G\to\infty$.}
\label{eq:etaEA_Ginf}
\ee
The challenging non-EA zero-quantum-capacity threshold $\eta>1/2$ is relaxed to the EA threshold $\eta>\kappa_E$ now, which is always achievable via overcoupling the cavity.

\begin{figure}
    \centering
    \includegraphics[width=\linewidth]{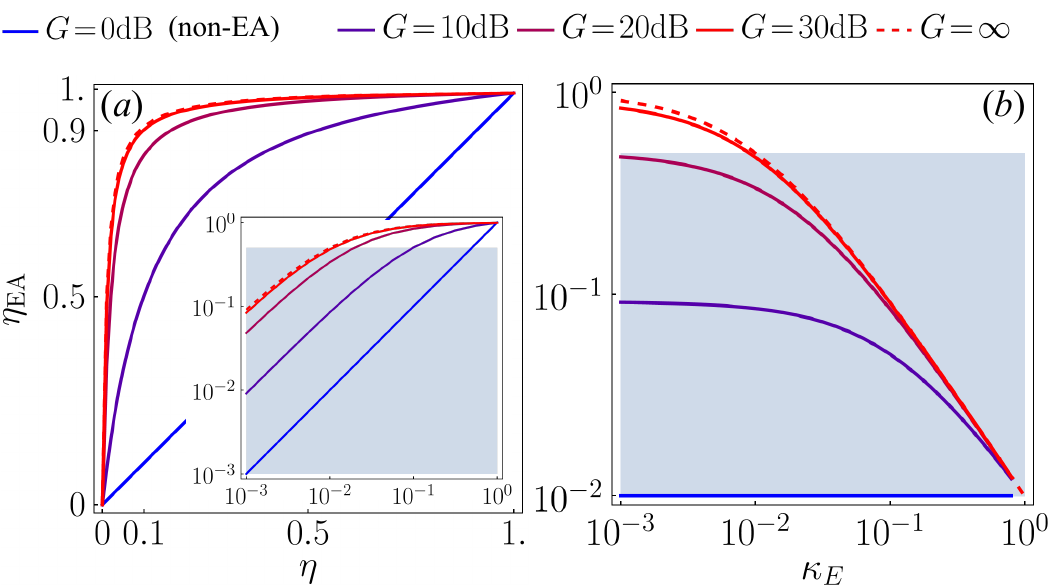}
    \caption{The EA transduction efficiency $\eta_{\rm EA}$ versus: (a) the non-EA efficiency $\eta$, with $\kappa_E=0.01$; (b) the intrinsic loss $\kappa_E$, with $\eta=0.01$. Inset in (a): Zoom-in near $\eta\to 0$ in logarithmic scale. Squeezer gain $G$: from blue to red, $G$ increases from 0dB (non-EA) to 30dB by step of 10dB. 
    Red dashed: $G\to\infty$ [\eqref{eq:etaEA_Ginf}]. Blue-shaded region: zero quantum-capacity region $\eta_{\rm EA}\le 1/2$. antisqueezer gain $G'$ is chosen according to \eqref{eq:optGp_nonideal}. }
    \label{fig:EAeffvsEta}
\end{figure}

We evaluate the EA advantage in transduction efficiency in Fig.~\ref{fig:EAeffvsEta}. In subplot (a), we fix the intrinsic loss $\kappa_E=0.01$ and vary the non-EA efficiency $\eta$. The EA efficiency overwhelms the non-EA efficiency (blue diagonal line), even with an intermediate-scale, near-term available squeezing gain $G=10$dB. In the inset, we observe the EA efficiency $\eta_{\rm EA}$ surpasses the zero-quantum-capacity threshold $1/2$ (upper boundary of blue-shaded region) at $\eta\gtrsim \kappa_E= 0.01$, given squeezing gain $G=30$dB---as predicted by \eqref{eq:etaEA_Ginf}.  
In subplot (b), we fix $\eta=0.01$ and vary $\kappa_E$. At high squeezing, the zero-quantum-capacity threshold $\eta_{\rm EA}=1/2$ can be achieved for $\kappa_E\lesssim \eta=0.01$. For the minimum squeezing requirement, we observe that at least $G=20$dB is required for $\eta_{\rm EA}\ge 1/2$ in the best case $\kappa_E\to 0$, which can be predicted by the \hw{lossless coupler} formula \eqref{eq:etaEA_ideal}.

%TC:ignore
%TC:endignore

\section{Full cavity model of coupling}
\label{sec:full_cavity_model}

Now we proceed to the full cavity model for the nonlinear couplers. Without loss of generality, here we present the formulation of cavity electro-optical coupling. 
{A similar formulation of cavity electro-optomechanical coupling~\cite{andrews2014bidirectional,higginbotham2018harnessing} is included in Appendix~\ref{app:full_cavity_model_EMO}, which holds for general bosonic nonlinear coupling with a mediating mode.} 

In cavity electro-optics, probe $P$ and signal $S$ are carried on optical/microwave cavity modes, associated with annihilation operators $\hat a_P$, $\hat a_S$ satisfying the commutation relation $[\hat{a}_X,\hat{a}_{X}^\dagger]=1$, where $X=P,S$. 
The quality of the cavities are characterized by the cavity external coupling rates and intrinsic loss rates $\gamma_{X,c}$ and $\gamma_{X,0}$. In this paper, we adopt the alternative characterization with the total linewidths $\Gamma_X\equiv \gamma_{X,c}+\gamma_{X,0}$ and the coupling ratios $\zeta_X\equiv\gamma_{X,c}/\Gamma_X$. \hw{Here we consider the perfectly red-detuned pump for simplicity~\cite{Tsang2010,Tsang2011}, e.g. for microwave-to-optical transduction the optical pump frequency is perfectly at $\omega_{\rm pump}=\omega_P-\omega_S$, where $\omega_P, \omega_S$ are the cavity resonance frequencies of optical probe and microwave signal respectively. We defer the general formulation allowing imperfect pump detuning in Appendix~\ref{app:full_cavity_model_EO}, which demonstrates that our protocol is robust against the detuning imperfections.}
In the frame rotating with the cavity resonance frequencies, assuming the rotating wave approximation, an electro-optics system in the red sideband pumping case can be described by the effective Hamiltonian~\cite{Tsang2010,Tsang2011}: 
\be 
\hat H_I=-\hbar g(\alpha^* \hat{a}_S^{\dagger}\hat{a}_P+ \alpha \hat{a}_S  \hat{a}_P^\dagger),
\label{H_I_EO}
\ee 
where $g$ is the electro-optic nonlinear coupling coefficient in the unit of Hz and $\alpha$ is the in-cavity pump amplitude. The interaction strength is typically characterized by the cooperativity $C=4|g\alpha|^2/\Gamma_S\Gamma_P$.

Solving the steady-state solution of the quantum Langevin equation~\cite{gardiner1985input,Tsang2011} for $\hat H_I$ in the Fourier domain, we obtain the broadband version of \eqref{eq:channel}. The probe transmissivity spectrum is
\bal
\sqrt{\kappa(\omega)}e^{i\theta_P(\omega)}=-1+\frac{2\zeta_P(1-2i\frac{\omega}{\Gamma_S})}{(1-2i\frac{\omega}{\Gamma_P})(1-2i\frac{\omega}{\Gamma_S})+C}
\label{eq:kappa}
\,,
\eal  
the signal-to-probe transduction efficiency spectrum is
\be 
\sqrt{\eta(\omega)}e^{i\theta_S(\omega)}=\frac{2i\sqrt{C } \sqrt{\zeta_P\zeta_S }}{(1-2i\frac{\omega}{\Gamma_P})(1-2i\frac{\omega}{\Gamma_S})+C }
\,.
\label{eq:eta}
\ee
\hw{As a reminder, here $\omega$ is in the frame rotating with the cavity resonance frequencies.}
The intrinsic loss spectrum can be obtained correspondingly as $\kappa_E(\omega)=1-\kappa(\omega)-\eta(\omega)$. It is worthwhile to note that the cavity is asymptotically lossless ($\kappa_E(\omega)\to 0$) at the cavity overcoupling limit ($\zeta_P,\zeta_S\to 1$). For weak nonlinear coupling $C\ll 1$, the peak conversion efficiency  $\eta(\omega=0)=\zeta_P\zeta_S\cdot 4C/(1+C)^2$; the half-power bandwidth of $\eta(\omega)$ is $B\simeq \min\{\Gamma_S,\Gamma_P\}$.

%TC:ignore
\begin{figure*}
    \centering
    \includegraphics[width=0.95\linewidth]{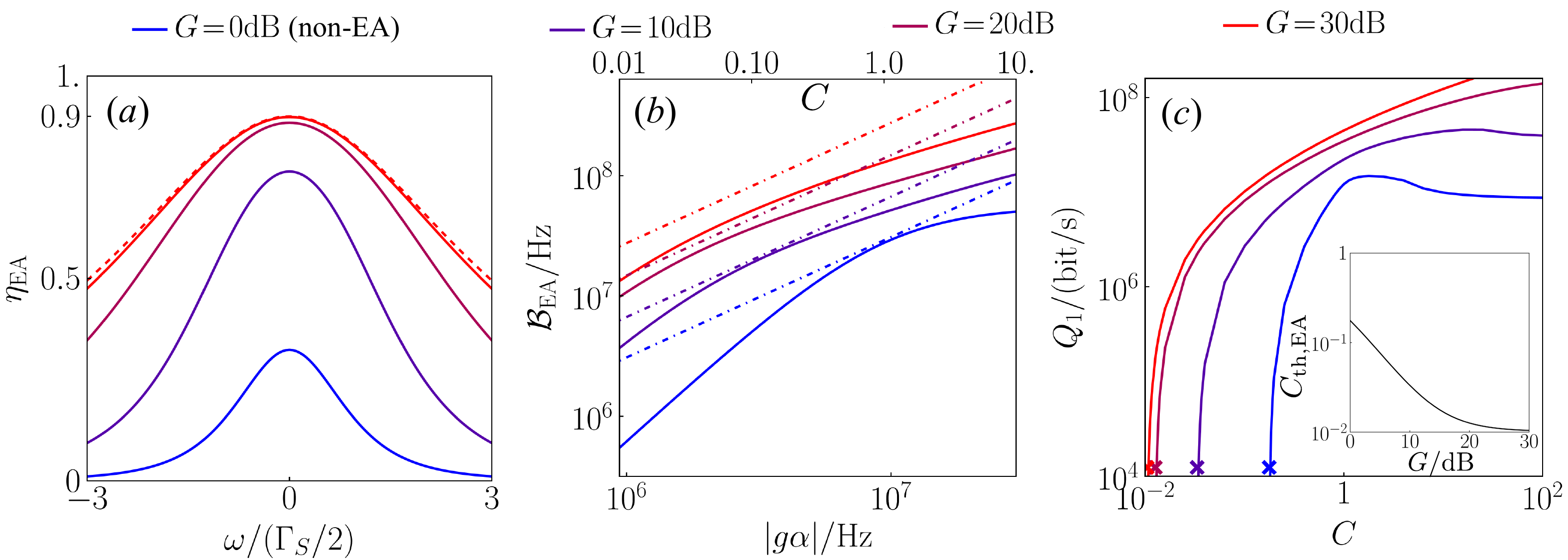}
    \caption{(a) EA transduction efficiency spectrum $\eta_{\rm EA}(\omega)$ under various squeezing gain $G$. Cooperativity $C=0.1$. The dashed line is the $G\to \infty$ limit obtained from \eqref{eta_EA_Ginfty}. (b) The EA efficiency-bandwidth product $\calB_{\rm EA}$ versus the effective nonlinear coupling strength $|g\alpha|$.
    The solid lines are under fixed $\Gamma_P$, $\Gamma_S$, for which we provide the cooperativity $C$ values as the upper axis ticks; while the dot-dashed lines are under optimized $\Gamma_P$, $\Gamma_S$ that maximize $\calB_{\rm EA}$ for a given $|g\alpha|$. 
    (c) Broadband quantum capacity rate $Q_1$ versus cooperativity $C$ under various squeezing gain $G$. The crosses indicate the zero-quantum-capacity thresholds $C_{\rm th}$ under each $G$ in \eqref{C_EA_threshold}. Inset: $C_{\rm th}$ versus $G$(dB), as given in \eqref{C_EA_threshold}. The $G=0$dB point goes back to \eqref{C_DC_requirement}.
    In all figures, $\zeta_P=\zeta_S=0.99$; Linewidths $\Gamma_P=25.8$MHz, $\Gamma_S=13.706$MHz are chosen according to the high-cooperativity setup in Ref.~\cite{sahu2022quantum}, except the dot-dash lines of (b).
     }
    \label{fig:etaEA_EBPEA}
\end{figure*}
%TC:endignore

Now we demonstrate the EA advantage. For simplicity, we assume a broadband two-mode squeezing for the squeezer $G(\omega)=G$; for the gain in the antisqueezing, however, the optimal choice of phase matching $\theta_P(\omega)$ and gain $G^{\prime\star}(\omega)$ in \eqref{eq:optGp_nonideal} will be frequency dependent.
Nevertheless, these requirements can be achieved by properly engineering the squeezing cavities, \QZ{e.g. using a sequential array of parametric amplifiers as we present in Appendix~\ref{app:freq_dependent_antisqz}}.
In particular, frequency-dependent squeezing is already being  utilized in Laser Interferometer Gravitational-Wave Observatory (LIGO)~\cite{mcculler2020frequency,jia2024squeezing}. According to \eqref{eq:etaEA_full}, the EA transduction efficiency is
$
\eta_{\rm EA}(\omega)= \eta\left(\omega\right) {G}/{\left[G \left(1-\kappa\left(\omega\right) \right)+\kappa\left(\omega\right) \right]}.
$
At the $G\to\infty$ limit, we obtain a closed-form expression
\bal 
\eta_{\rm EA}(\omega)|_{G\to \infty}=\frac{C \Gamma _S^2 \zeta _S}{\Gamma _S^2 \left(C+1-\zeta _P\right)+4 \omega ^2
   \left(1-\zeta _P\right)}.
   \label{eta_EA_Ginfty}
\eal 
With entanglement assistance, we observe an improvement in the peak efficiency $\eta_{\rm EA}(\omega=0)|_{G\to \infty}={\zeta_S}/{\left[1+(1-\zeta _P)/C\right]}$, and a bandwidth broadening $B_{\rm EA}|_{G\to \infty}=\sqrt{1+C/(1-\zeta_P)}\Gamma_S$. Remarkably, the EA bandwidth no longer depends on the probe linewidth $\Gamma_P$ at $G\to \infty$. Hence the EA advantage is not limited to the weak nonlinear coupling scenarios: even though the on-resonance efficiency can get close to unity with stronger pumping merely, entanglement allows broadband improvement via bandwidth broadening. Similar quantum advantages using non-classical probes have been found in cavity dark matter searches~\cite{malnou2019squeezed,shi2023ultimate}.

We plot an example of the EA conversion efficiency spectrum $\eta_{\rm EA}(\omega)$ in Fig.~\ref{fig:etaEA_EBPEA}(a). As predicted, we see that the bandwidth of the non-EA case ($G=0$dB) is approximately $\Gamma_S$, and the EA bandwidth grows as $G$ increases in addition to the peak efficiency advantage. 

Below, we quantify the EA advantage with three measures of transduction performance: efficiency-bandwidth product, minimum threshold of cooperativity for quantum communication, and broadband quantum capacity.

{
\subsection{ Fundamental limit on efficiency-bandwidth product}}
\label{sec:fundamental_EBP}

To quantify the broadband transduction efficiency, we define efficiency-bandwidth product (EBP) as the integration of transduction efficiency over the entire spectrum (see \eqref{def_EBP}). This metric is particularly useful for broadband quantum sensing applications~\cite{shi2023ultimate}.
\begin{theorem}
\label{theorem:EO}
(EBP limit for electro-optical transduction)
Without entanglement assistance, the EBP of quantum transduction with Hamiltonian \eqref{H_I_EO} \hw{{\rm [}which leads to lineshape $\eta(\omega)$ \eqref{eq:eta}{\rm ]}} is
\bal 
\calB &\equiv \int_{-\infty}^\infty\eta(\omega) d\omega =\frac{2 \pi  C\Gamma_P\Gamma_S \zeta_P \zeta_S}{(C+1)(\Gamma_P+\Gamma_S)}\le \calB_{\rm max}\,,
\eal
which achieves the maximum
$
\calB_{\rm max}\equiv \pi\zeta_S\zeta_P|g\alpha|\le \pi |g\alpha|,
$ 
at ${\Gamma}_P={\Gamma}_S=2|g\alpha|$ (i.e. $C=1$) given a fixed $|g\alpha|$.
\end{theorem}
The full derivation is in Appendix~\ref{app:EBP_EO}. $\calB_{\rm max}$ is a fundamental limit for the non-EA case determined by the nonlinear coupling coefficient $g$ and pump power $\propto |\alpha|^2$, which is independent on any higher-$Q$ cavity engineering. {For electro-optomechanical transducers, we present a similar limit in Subsection~\ref{sec:EMO}.}

Meanwhile, we also derive a closed-form expression of EA EBP $\calB_{\rm EA} \equiv \int_{-\infty}^\infty\eta_{\rm EA}(\omega) d\omega$, which is too lengthy to be displayed here. Under the same cavity setup $\Gamma_P=\Gamma_S=2|g\alpha|$, $\zeta_P=\zeta_S=1$, we obtain $\calB_{\rm EA} =G^{1/4}\pi |g\alpha|\ge G^{1/4}\calB_{\rm max}$, breaking the fundamental limit $\calB_{\rm max}$ of the non-EA case. Allowing freely choosing ${\Gamma}_P={\Gamma}_S=2\sqrt{G}|g\alpha|$, we have $\calB_{\rm EA}\simeq 0.703 \sqrt{G} \pi |g\alpha|$ with $\sqrt{G}$ advantage compared to $\calB_{\rm max}$. 
When $G\to \infty$,
\be 
\calB_{\rm EA}|_{G\to \infty}\to \frac{\pi C\Gamma_S\zeta_S}{2\sqrt{(1+C-\zeta_P)(1-\zeta_P)}},
\label{B_EA_infty}
\ee 
which diverges as $\zeta_P\to 1$ as expected.

While the above optimal results provide the ultimate limits, here we also consider the practical case of low cooperativity $C\ll 1$. In this case, entanglement can enhance EBP by a factor of $G$,
$
\calB_{\rm EA}=G\cdot \calB\,
$
when $\zeta_P=\zeta_S=1$.

Under imperfect $\zeta_P,\zeta_S<1$, we plot $\calB_{\rm EA}$ in Fig.~\ref{fig:etaEA_EBPEA}(b). Still, we observe orders of magnitude of EA advantage. Either with $\Gamma_S,\Gamma_P$ fixed (solid lines) or with optimized $\Gamma_S,\Gamma_P$ over each given $|g\alpha|$ (dot-dashed lines), the EA advantages are demonstrated over the maximal non-EA EBP $\calB_{\rm max}$ (blue dot-dashed line) over a wide range of effective nonlinear coupling strength $|g\alpha|$, corresponding to cooperativity $C\in [0.01,10]$ for the $\Gamma_S,\Gamma_P$ fixed cases. 

% [The advantage of dashed lines here (Gamma optimized) only scales ~1/G^1/4, different from the sqrt G prediction, since the optimal Gamma also ~G^1/4, different from prediction. This is because here zeta is not exacly 1, despite 0.99 very close to 1 (very sensitive to zeta). I have checked for zeta=1, advantage ~ sqrt G and optimal Gamma~sqrt G. ]

\subsection{Threshold of cooperativity and broadband quantum information rate}
\label{sec:broadband_Q}

While the efficiency-bandwidth product provides an intuitive characterization of the transduction efficiency, the ultimate quantum information transmission rates are characterized by the quantum capacity~\cite{lloyd1997capacity,shor2002quantum,devetak2005private} across the entire spectrum. When the environment are cooled to vacuum, the one-way quantum capacity of transducer is given by \eqref{def_Q}~\cite{holevo2001evaluating,wang2022quantum}.

At the weak nonlinear coupling limit, the maximum transduction efficiency locates at the on-resonance frequency $\omega=0$. For the non-EA case, we have
$
\eta(0)={4 C\zeta_P\zeta_S}/{(1+C)^2}
$
which surpasses the zero-capacity threshold $1/2$ only when~\cite{wu2021deterministic}
\be 
C\ge C_{\rm th}=  -1+4\zeta_{ S} \zeta_{ P}-\sqrt{8\zeta_{ S} \zeta_{ P} (2 \zeta_{ S} \zeta_{ P}-1)}\ge 3-2\sqrt{2}.
\label{C_DC_requirement}
\ee

With the EA boost, we have the threshold
\bal 
C_{\rm th,EA}= &-1+\zeta_P \left(\left(4 \zeta_S-2\right)
   G+2\right)
   \\
&   -2 \sqrt{\zeta_P G
   \left(\zeta_P \left(4
   \zeta_S+(1-2 \zeta_S)^2
   G-1\right)-2 \zeta_S\right)}.
\label{C_EA_threshold}
\eal 
When $G\to\infty$, the threshold converges to $C_{\rm th}\to (1-\zeta_P)/(2\zeta_S-1)$ when $\zeta_S\ge 1/2$.

Additional insight can be obtained by considering the overcoupling limit of $\zeta_P=\zeta_S=1$, threshold in \eqref{C_EA_threshold} leads to 
$ 
C_{\rm th,EA}|_{\zeta_P=\zeta_S=1}= 1/{(\sqrt{G}+\sqrt{1+G})^2},
$
which is lowered by a factor of $1/G$ asymptotically. It is easy to check that $\eta_{\rm EA}(0)$ approaches unity at the large $G$ limit. We plot the threshold in the inset of Fig.~\ref{fig:etaEA_EBPEA}(c) for a practical case and identify a reduction by over an order of magnitude when squeezing gain $G$ is large.

%The threshold of cooperativity to enable a non-zero rate is reduced by a factor $\propto G$, as we show in Appendix. 
We plot $Q_1$ versus the cooperativity $C$ for different gain $G$ in Fig.~\ref{fig:etaEA_EBPEA}(c). 
Merely $G=10$dB squeezing is sufficient to enable orders of magnitude advantage at low cooperativity. Remarkably, for large $C$ the quantum capacity without probe-ancilla entanglement begins to decay with $C$ as the cavity goes into the oscillatory region with Rabi splitting; in contrast, the quantum capacity assisted by probe-ancilla entanglement can further increase with $C\gg 1$.

\subsection{Generalization to transduction with intermediate modes} 
\label{sec:EMO}

Microwave-optical quantum transduction is known to be enhanced by mediating modes. As an example, we focus on the electro-optomechanical transduction, which yields the state-of-the-art efficiency so far ~\cite{brubaker2022optomechanical}.
In the frame rotating at the cavity resonance frequencies for microwave and optical modes, the cavity electro-optomechanical dynamics can be described by the effective Hamiltonian~\cite{gardiner1985input,bowen2015quantum,andrews2014bidirectional}
\be 
H_I=
% \\&\quad 
\hbar g_S \hat a_S^\dagger \hat a_S \hat x_M+ \hbar g_P \hat a_P^\dagger \hat a_P \hat x_M
\label{H_I_EMO}
\ee
where $\hat a_S,\hat a_P,\hat a_M$ are the annihilation operators of the signal (microwave/optical), probe (optical/microwave), and mediating (mechanical) modes, and $\hat x_M=x_{\rm zp}(\hat a_M +\hat a_M^\dagger)$ with $x_{\rm zp}=\sqrt{\frac{\hbar}{2m\omega_M}}$ {being the zero-point motion}. {The mechanical oscillator has mass $m$ and frequency $\omega_M$.} Here the nonlinear coupling coefficients $g_S,g_P$ are in unit of $\rm Hz\cdot m^{-1}$. We define $\calG_{S,P}\equiv g_{S,P}x_{\rm zp}|\alpha_{S,P}|$ proportional to the nonlinear coupling coefficients and pumping amplitudes, analogous to $g|\alpha|$ in the electro-optical coupling~\cite{Tsang2011}. 

Consider red sideband pumping and the resolved sideband limit, the electro-optomechanical coupling yields the beamsplitter-type input-output relation similar to the electro-optics, up to a different spectral lineshape. Thus most of our conclusions for the electro-optics can be trivially generalized to the electro-optomechanics. Here we present the EBP limit.
\begin{theorem}
\label{theorem:EMO}
(EBP limit for electro-optomechanical transduction)
Without entanglement assistance, the EBP of quantum transduction with Hamiltonian \eqref{H_I_EMO} is upper bounded by \eqref{eq:Bmax_EMO_app} as a function of $\calG_P,\calG_S$. 

Specifically, in the symmetric case of ${\calG_P}={\calG_S}=\calG$, the EBP is upper bounded by
\bal 
\calB_{{\calG_P}={\calG_S}}\le \frac{\sqrt{107+51\sqrt{17}}}{32}\pi \zeta_P\zeta_S \calG \simeq 1.749 \zeta_P\zeta_S \calG.
\eal
%achieves the maximum at ${\Gamma}_P={\Gamma}_S=\sqrt{2(3+\sqrt{17})}{\calG_P}\simeq 3.774{\calG_P}$ given fixed ${\calG_P},{\calG_S}$.
\end{theorem}
The full derivation is in Appendix~\ref{app:EBP_EMO}.

Our EA transduction protocol also applies to electro-optomechanical transduction. Similar enhancement in EBP proportional to two-mode squeezing gain $G$ can overcome the above fundamental EBP limit.  {At the overcoupling limit $\zeta_P,\zeta_S\to 1$ and lossless mechanical resonator, the overall coupling loss $\kappa_E(\omega)\to 0$, then the signal can always be perfectly recovered with strong squeezing $G\to \infty$ as $\eta_{\rm EA}(\omega)|_{\kappa_E\to 0, G\to \infty}\to 1$ .}

%\section{ Experiment design---}
%{provide the system design here with experimental pieces}

\section{ Discussions} 
\label{sec:discussions}
The most challenging part of the proposed EA transduction is the frequency-dependent inline squeezing. In the optical-to-microwave transduction, the required microwave inline squeezing can be readily realized to high gain~\cite{backes2021quantum,xu2023magnetic,qiu2023broadband}. Alternative to realizing optical inline squeezing for microwave-to-optical transduction, one can also utilize the optical-to-microwave transduction to generate optical-microwave entanglement from optical-optical entanglement~\cite{vahlbruch2016detection,kashiwazaki2020continuous}, then teleportation enables bi-directional transduction~\cite{zhong2020proposal,zhong2022microwave,wu2021deterministic}. \hw{We present a simple frequency-dependent antisqueezer design using a sequential array of cavity parametric amplifiers in Appendix~\ref{app:freq_dependent_antisqz}, which approaches the desired antisqueezing spectrum well and achieves scalable advantage in the broadband quantum capacity over the non-EA transduction.}

{In this paper we have focused on the beamsplitter-type nonlinear couplers. We expect the intraband entanglement to similarly enhance  the squeezing-type couplers, which we leave to future study. We note that the squeezing-type transducers cannot perform any quantum transduction without our proposal of intraband entanglement assistance, because it forms a phase-conjugate amplifier of zero quantum capacity~\cite{rau2022entanglement,holevo2008entanglement}.}

%are effectively phase-conjugate amplifiers, which 
% invoke quantum noises proportional to the gain~\cite{caves1982quantum} and 
%are entanglement-breaking~\cite{holevo2008entanglement}, thus forbidding any quantum transduction due to zero quantum capacity~\cite{rau2022entanglement}. Therefore, we have beamsplitter-type to } 

% Finally, we compare our proposal to the existing works.
Compared with the proposal with in-cavity squeezing~\cite{zhong2022quantum} that boosts a single quadrature transduction, our approach allows transduction of both quadratures thus is free from encoding in the ideal case, and does not require additional pumping at the cavity that can lead to additional heating. \hw{We note that the in-cavity squeezing protocol~\cite{zhong2022quantum} requires the cavity system to be on resonance with minimal detuning, resulting in a highly limited operating bandwidth. In contrast, our protocol is robust against the detunings (as shown in Section~\ref{sec:full_cavity_model} and Appendix~\ref{app:eval_pumpdetuning}) and is only subject to frequency-independent requirements of relatively low intrinsic loss (achievable by overcoupled cavities) and strong squeezing, thus enabling broadband transduction.} Moreover, an explicit protocol that recovers the initial quantum state is absent in Ref.~\cite{zhong2022quantum}. Compared with the GKP-based protocol in Ref.~\cite{wang2024passive} that requires the input signal to be GKP encoded, our protocol relies on less challenging quantum resources of inline squeezing. Distinct from both the two protocols above, our proposal lifts the requirement of encoding and thus can be applied to transduce general bosonic quantum states more compatible with existing optical communication infrastructures. While Ref.~\cite{wang2024passive} only considers the perfect cavity of $\kappa_E= 0$, our protocol shows advantage for general scenarios. Compared with the adaptive protocol~\cite{zhang2018quantum}, our protocol does not require the precise broadband homodyne measurement and adaptive control which include delay lines that increase the loss and limit the capacity and speed of transduction. We note that our proposal requires two-mode squeezers, similar to the single-mode squeezers in Ref.~\cite{zhang2018quantum}, of which the bandwidth are being actively increased~\cite{qiu2023broadband,nehra2022few,yang2021squeezed}.

\begin{acknowledgements}

HS and QZ proposed the protocol in discussion, performed analyses, generated the figures and wrote the manuscript. The project is supported by Office of Naval Research Grant No. N00014-23-1-2296 and National Science Foundation Engineering Research Center for Quantum Networks Grant No. 1941583. QZ also acknowledges support from DARPA MeasQUIT HR0011-24-9-0362, National Science Foundation OMA-2326746, National Science Foundation CAREER Award CCF-2240641 and an unrestricted gift from Google. This material is partially based upon work supported by the U.S. Department of Energy, Office of Science, National Quantum Information Science Research Centers, Superconducting Quantum Materials and Systems Center (SQMS) under the contract No. DE-AC02-07CH11359.

\end{acknowledgements}

% \bmsection{Supplemental document}
% See Supplement 1 for supporting content. 

\begin{widetext}

\begin{appendix}
\section*{Appendix}
% \noindent{\bf Appendix on backgrounds}\\
% The need of transduction comes from the trade-off between photon transmission loss and strength of nonlinearity at different frequencies. While optical photons are ideal for transmitting quantum information over long distances due to low loss and low noise at room temperature, the optical nonlinearity at optical frequencies is much weaker than that achievable at microwave---the frequency where major quantum information processors operate at. In addition, physical-layer data at different frequencies are of importance in sensing applications while quantum resources such as squeezing are more challenging at lower frequencies such as kilohertz. 
% On-chip system typically have overall transduction efficiency on the order of $\sim 10^{-5}$ across a bandwidth of 10MHz~\cite{holzgrafe2020cavity,mirhosseini2020superconducting}.   
% In the electro-optomechanical platforms, on-resonance efficiency can be pushed up to 47\%~\cite{brubaker2022optomechanical} at the cost of bandwidth being narrowed down to $\sim 2$kHz. Alternative approaches adopt a pulsed pump with high power to achieve high efficiency; however, such an approach is not compatible with general continuous-wave signals, and the pulsed pump will lead to heating that significantly reduces the repetition rate of transduction to $10\sim100$Hz~\cite{sahu2022quantum,qiu2023coherent,sahu2023entangling}, far from what is needed to satisfy quantum networking. 

\subsection{Full derivation of the overall input-output relation of entanglement-assisted transducer}
\label{app:input-output}
Consider initial probe and ancilla modes $\hat \calE_{P_0}, \hat \calE_{A_0}$ in vacuum.
The two-mode squeezer before nonlinear coupling gives
\bal 
\hat \calE_P=\sqrt{G} \hat \calE_{P_0}+\sqrt{G-1} \hat \calE_{A_0}^{\dagger }\,,\\
\hat \calE_A=\sqrt{G-1} \hat \calE_{P_0}^{\dagger }+\sqrt{G} \hat \calE_{A_0}\,.
\eal 
The nonlinear coupling forms a beamsplitter between the squeezed probe $\hat \calE_{P}$ and the signal $\hat \calE_{S}$:
\bal 
\hat \calE_{P'}=e^{i\theta_P}\sqrt{\kappa} \hat \calE_P+e^{i\theta_S}\sqrt{\eta} \hat \calE_{S}+\sqrt{\kappa_E}\hat \calE_E,
\label{eq:coupling_app}
\eal 
while ancilla $\hat\calE_{A}$ is intact. Here, the intrinsic loss $\kappa_E=1-\eta-\kappa$ and the environment mode $\hat \calE_{E}$ is in vacuum.
After the nonlinear coupling, the antisqueezer, with phase compensation $-\theta_P$ on probe, gives the final output
\ba 
&\hat \calE_{P^{\rm out}}&=e^{-i\theta_P}\sqrt{G'} \hat \calE_{P'}-\sqrt{G'-1} \hat \calE_{A}^{\dagger }
\nonumber\\
&&=e^{-i\theta_P}\sqrt{G'} \left(e^{i\theta_P}\sqrt{\kappa} \hat \calE_P+e^{i\theta_S}\sqrt{\eta} \hat \calE_{S}+\sqrt{\kappa_E}\hat \calE_E\right)
\nonumber\\
&&\quad-\sqrt{G'-1} \hat \calE_{A}^{\dagger }
\nonumber\\
&&= \left(\sqrt{G \kappa
     G'}-\sqrt{(G'-1)(G-1)}\right) \hat \calE_{P_0}
\nonumber\\
&&\quad +e^{i(\theta_S-\theta_P)}\sqrt{\eta  G'} \hat \calE_S \nonumber
\nonumber\\
&&\quad+
\left(\sqrt{(G-1) \kappa  G'}   - \sqrt{(G'-1)G} \right)
   \hat \calE_{A_0}^{\dagger }
\nonumber\\
&&\quad+
    e^{-i\theta_P}\sqrt{\left(1-\eta -\kappa \right) G'}\hat \calE_E \,.
\label{eq:outputprobe_general_app}
\ea
To keep the output to vacuum when the input signal is vacuum, one needs to annihilate the coefficient in front of $\hat \calE_{A_0}^{\dagger }$ in \eqref{eq:outputprobe_general_app}, leading to 
\be 
G'\gets G^{\prime\star}\equiv \frac{1}{1-\kappa+\kappa/G}.
\ee
By such antisqueezing, finally the output reduces to 
\bal 
&\hat \calE_{P^{\rm out}}^\star=
\\
&\frac{\sqrt{\kappa }\hat \calE_{P_0}+ \sqrt{\eta  G} e^{i(\theta_S-\theta_P)}\hat \calE_S +  \sqrt{(1-\eta-\kappa) G} e^{-i\theta_P}\hat \calE_E}{\sqrt{G \left(1-\kappa \right)+\kappa}}\,.
\eal
Note that $\hat \calE_{P_0}, \hat \calE_E$ are in vacuum state, the output can be written as
\be 
\hat \calE_{P^{\rm out}}^\star=
 \sqrt{\eta_{\rm EA}} e^{i(\theta_S-\theta_P)}\hat \calE_S + \sqrt{1-\eta_{\rm EA}}\hat \calE_{\rm VAC} \,,
\ee
where the noise background $\hat \calE_{\rm VAC}$ is in vacuum state, we define the entanglement-assisted (EA) transduction efficiency as
\be 
\eta_{\rm EA} \equiv  \frac{\eta G}{G (1-\kappa )+\kappa }\,.
\ee

\hw{At the same time, we can obtain the ancilla output 
\ba 
&\hat \calE_{A^{\rm out}}&=-e^{i\theta_P}\sqrt{G'-1} \hat \calE_{P'}^\dagger+\sqrt{G'} \hat \calE_{A}
\nonumber\\
&&=-e^{i\theta_P}\sqrt{G'-1} \left(e^{-i\theta_P}\sqrt{\kappa} \hat \calE_P^\dagger+e^{-i\theta_S}\sqrt{\eta} \hat \calE_{S}^\dagger+\sqrt{\kappa_E}\hat \calE_E^\dagger\right)
\nonumber\\
&&\quad+\sqrt{G'} \hat \calE_{A}
\nonumber\\
&&= \left(-\sqrt{G \kappa
     (G'-1)}+\sqrt{G'(G-1)}\right) \hat \calE_{P_0}^\dagger
\nonumber\\
&&\quad -e^{i(\theta_P-\theta_S)}\sqrt{\eta  (G'-1)} \hat \calE_S^\dagger \nonumber
\nonumber\\
&&\quad+
\left(-\sqrt{(G-1) \kappa  (G'-1)}   + \sqrt{G'G} \right)
   \hat \calE_{A_0}
\nonumber\\
&&\quad-
    e^{-i\theta_P}\sqrt{\left(1-\eta -\kappa \right) (G'-1)}\hat \calE_E \,
\\
&&
\!\!\!\!\!\stackrel{ G'\gets G^{\prime\star}}{=}
-e^{i(\theta_P-\theta_S)}\sqrt{\eta_{\rm EA}\kappa (1-1/G)} \hat \calE_S^\dagger
\nonumber
\\
&&\quad 
+(1-\kappa)\sqrt{\frac{G-1}{1-\kappa+\kappa/G}}  \hat \calE_{P_0}^\dagger+{\rm vac}
\label{eq:output_ancilla_general_app}
\ea
where vac represents the vacuum noise terms. The ancilla output is dominated by the quantum amplification noise at the limit $G\to \infty$ with $\eta>0$.
}

\ 

\ 

\  

\ 

\ 

\ 

\ 

\ 

\

\subsection{Impact of loss in imperfect squeezing}
\label{app:ancilla_loss}
Now we investigate the impact of loss in imperfect squeezers and antisqueezers. In fact, we only need to consider the loss for the ancilla after the squeezer $\calS(G)$ and before the antisqueezer $\calS^\dagger (G')$. This is because any loss before the squeezer does not change the vacuum state of input ancilla and any loss after the antisqueezer does not affect the output probe (we discard the output ancilla since it is dominated by amplification noise when $G$ is large), meanwhile any additional signal loss between squeezer and antisqueezer can be merged into the intrinsic loss $\kappa_E$, which is already included in the model of the main text. Therefore, one can also regard the squeezing related loss as imperfect ancilla storage.

Denote the ancilla storage efficiency as $\kappa_A$ (with loss $1-\kappa_A$), \eqref{eq:coupling_app} becomes
\bal 
\hat \calE_{P'}&=e^{i\theta_P}\sqrt{\kappa} \hat \calE_P+e^{i\theta_S}\sqrt{\eta} \hat \calE_{S}+\sqrt{\kappa_E}\hat \calE_E\,,\\
\hat \calE_{A'}&=\sqrt{\kappa_A} \hat \calE_A+\sqrt{1-\kappa_A} \hat \calE_{F}\,,
\eal 
where $F$ is the environment mode involved in the ancilla loss, initialized in vacuum state.
With the additional loss, the final output probe in \eqref{eq:outputprobe_general_app} becomes 
\begin{align}
\hat \calE_{P^{\rm out}}&=e^{-i\theta_P}\sqrt{G'} \hat \calE_{P'}-\sqrt{G'-1} \hat \calE_{A'}^{\dagger }=\left(\sqrt{G \kappa
     G'}-\sqrt{(G'-1)(G-1)\kappa_A}\right) \hat \calE_{P_0} \nonumber
% \nonumber\\
% &&\quad 
+e^{i(\theta_S-\theta_P)}\sqrt{\eta  G'} \hat \calE_S \nonumber
% \nonumber\\
% &&\quad
+
\\
&\quad\left(\sqrt{(G-1) \kappa  G'}   - \sqrt{(G'-1)G\kappa_A} \right)
   \hat \calE_{A_0}^{\dagger }
% \nonumber\\
% &&\quad
+
    e^{-i\theta_P}\sqrt{\left(1-\eta -\kappa \right) G'}\hat \calE_E 
% \nonumber\\
% &&\quad 
-\sqrt{(G'-1)(1-\kappa _A)} \hat\calE_F^{\dagger }.
\end{align}
The last term invokes an additional amplification noise background, which leads to the total noise background of thermal photon number
\be 
N_B=\left(\sqrt{(G-1) G' \kappa }-\sqrt{(G'-1)G \kappa _A} \right){}^2+(G'-1) \left(1-\kappa _A\right)
\label{eq:Nb_ancillaloss}
\ee 
as a consequence of the nonzero ancilla loss $1-\kappa_A$.
In this case, the thermal noise in the output probe cannot be fully cancelled any more. The optimal choice that minimizes the output noise is 
\be 
G^{\prime\star}=\frac{(G-1) \kappa _A+\sqrt{(G-1)^2 \kappa _A^2-2 (G-1) \kappa _A (G \kappa +\kappa -1)+((G-1) \kappa +1)^2}+(G-1) \kappa +1}{2 \sqrt{2 \kappa  \left(G^2 \left(-\kappa _A\right)+\kappa _A+G-1\right)+\left((G-1)
   \kappa _A+1\right){}^2+(G-1)^2 \kappa ^2}}.
\ee 
We note that the additional noise \eqref{eq:Nb_ancillaloss} does not increase with gain $G$. Thus, at the strong squeezing regime $G\gg 1$, the optimal choice of $G'$ is similar to \eqref{eq:optGp_nonideal} as
\be
G^{\prime\star}\simeq \frac{G \kappa _A}{G (\kappa _A- \kappa) +\kappa }\simeq \frac{\kappa _A}{\kappa _A- \kappa }\,.
\ee
Accordingly, the EA efficiency $\eta_{\rm EA}$ is similar to \eqref{eq:etaEA_full} as
\be 
\eta_{\rm EA}= \eta G^{\prime\star} \simeq \eta \frac{G\kappa_A}{G (\kappa _A- \kappa)+\kappa }\simeq \eta \frac{\kappa_A}{\kappa _A- \kappa }\,.
\ee
In the presence of noise, the quantum capacity formula in \eqref{def_Q} does not apply. Instead, we adopt a lower bound of the quantum capacity (in the unit of bits) of the resulting bosonic thermal loss channel of transmissivity $\eta_{\rm EA}$ and additive thermal background photon number $N_B$ is~\cite{holevo2001evaluating}
\be 
q_{\rm LB}=\max\left\{0, \log_2 \left(\frac{\eta_{\rm EA}}{\left| 1-\eta_{\rm EA}\right| }\right)-g\left(\frac{N_B}{\left|
   1-\eta_{\rm EA}\right| }\right)
  \right\},
\label{eq:Q_LB_app}
\ee
where $g(x)\equiv (x+1)\log_2 (x+1)-x\log_2 x$.

Now we evaluate the impact of the ancilla loss on the quantum communication rate of the EA transduction. 
As shown in Fig.~\ref{fig:qVsEta_ancillaloss}, the advantage of our protocol is robust to ancilla loss. The advantage survives losses as severe as $10^{-1}=10\%$. Considering that the ancilla is not involved in the nonlinear signal-probe coupling which invokes extra losses, we expect $1-\kappa_A$ can be typically maintained lower than $\kappa_E$ which is $1\%$ here. Such robustness comes from the fact that the probe encounters loss $1-\kappa=\kappa_E+\eta$ during the nonlinear coupling process, and therefore we expect when $1-\kappa_A<\kappa_E+\eta$, the ancilla can enhance transduction as a reference with smaller loss. 

\begin{figure}
{    
\centering
\includegraphics[width=0.6\linewidth]{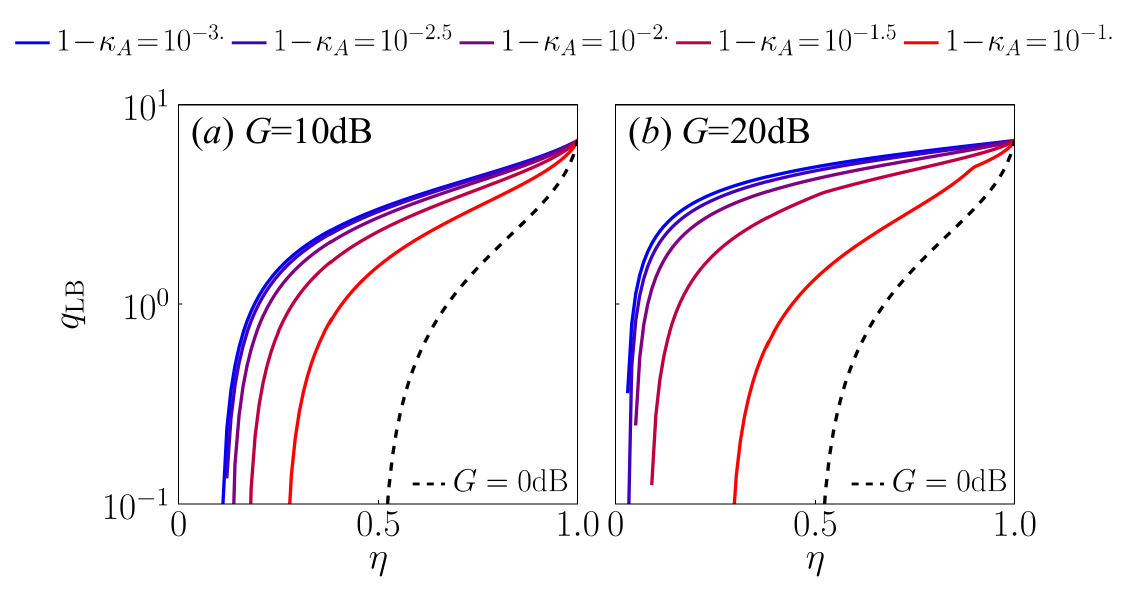}
    \caption{\hw{Quantum capacity lower bound $q_{\rm LB}$ versus non-EA transduction efficiency $\eta$ under various ancilla loss $1-\kappa_A$. (a)$G=10$dB; (b) $G=20$dB. Black dashed: non-EA, $G=0$dB. Intrinsic loss $\kappa_E=0.01$.}}
    \label{fig:qVsEta_ancillaloss}
    }
\end{figure}

%we observe that the ancilla loss $1-\kappa_A$ plays a role similar to the signal-probe coupling loss $\kappa_E$.

%The degradation of the performance due to $1-\kappa_A$ is insignificant as long as $1-\kappa_A<\kappa_E$. This is because $G'\simeq 1/\eta$ is finite even at $G\to \infty$, thus the strength of the ancilla-loss-invoked noise term $-\sqrt{(G'-1)(1-\kappa _A)} \hat\calE_E^{A\dagger }$ is finite $\propto \sqrt{\frac{1-\kappa_A}{\kappa_E}}$ relative to the vacuum noise due to signal-probe coupling loss, which is negligible when $1-\kappa_A\ll \kappa_E$. Reducing $1-\kappa_A$ below $\kappa_E$ is easy since the ancilla does not need to be involved in the nonlinear signal-probe coupling. For $1-\kappa_A>\kappa_E$, the performance degradation becomes significant, while the EA protocol still achieves non-zero quantum capacity at $\eta<0.5$ where the non-EA protocols have zero quantum capacity.

\color{black}

%\section{ Formula of cooperativity threshold}

%\section{ Appendix on related works}

\subsection{Full derivation of EBP of electro-optical transduction}
\label{app:EBP_EO}
To solve the EBP, we make use of the integration formula
\be 
\int_{-\infty}^\infty d\omega \frac{1}{C_1+4\omega^2C_2+16\omega^4}= \frac{\pi }{\sqrt{2}
   \sqrt{C_1}
   \sqrt{C_2+2\sqrt{C_1}}}.
   \label{eq:integration_f}
\ee 
Without the EA [see \eqref{eq:eta}], we have
\be 
\eta(\omega)=\frac{4 C \Gamma_P^2
   \Gamma_S^2 \zeta_P
   \zeta_S}{(C+1)^2
   \Gamma_P^2 \Gamma_S^2+4 \omega^2 \left(-2 C
   \Gamma_P \Gamma_S+\Gamma_P^2+\Gamma_S^2\right)+16
   \omega^4}.
\ee
With EA and optimal $G^\prime$ [see \eqref{eq:etaEA_full}], we have
\bal
&\eta_{\rm EA}(\omega)=
\\
&\frac{4 C \Gamma_P^2
   \Gamma_S^2 \zeta_P
   \zeta_S G}{\Gamma_P^2 \Gamma_S^2 \left(4
   (C+1) \zeta_P
   (G-1)+(C+1)^2-4 \zeta_P^2 (G-1)\right)+4 \omega^2 \left(-2
   C \Gamma_P
   \Gamma_S+\Gamma_S^2+\Gamma_P^2 (4
   \zeta_P (\zeta_P-\zeta_P
   G+G-1)+1)\right)+16 \omega^4}.
\eal  
At the overcoupling limit of $\zeta_P=\zeta_S=1$, the above formula simplifies to
\be 
\eta_{\rm EA}(0)|_{\zeta_P=\zeta_S=1}=\eta(0)\frac{G}{[1+4(G-1)C/(C+1)^2]}.
\ee 

Now consider cooperativity $C=4|g\alpha|^2/\Gamma_S\Gamma_P$, we fix the pump power and the nonlinear coupling coefficient, and consider EBP as a function of the cavity parameters $\Gamma_S,\Gamma_P$. With \eqref{eq:integration_f}, we have without EA
\bal 
\calB &\equiv \int_{-\infty}^\infty\eta(\omega) d\omega =\frac{2 \pi  C\Gamma_P\Gamma_S \zeta_P \zeta_S}{C \Gamma_P+C \Gamma_S+\Gamma_P+\Gamma_S}
=8\pi |g\alpha|\zeta_P \zeta_S \frac{\tilde{\Gamma}_P\tilde{\Gamma}_S}{(\tilde{\Gamma}_P+\tilde{\Gamma}_S)(4+\tilde{\Gamma}_P\tilde{\Gamma}_S)},
\label{eq:EBP_noEA_app}
\eal
where we have defined $\tilde{\Gamma}_X=\Gamma_X/|g\alpha|$. 

For the EA formula, we can also obtain lengthy closed-form solution of EA EBP $\calB_{\rm EA} \equiv \int_{-\infty}^\infty\eta_{\rm EA}(\omega) d\omega$ from \eqref{eq:integration_f}, which we will not display here. For $\zeta_P=\zeta_S=1$, we have a slightly simpler result,
\be 
\calB_{\rm EA}=\frac{8 \pi  \tilde\Gamma_{P}
   \tilde\Gamma_{S} G
   |g\alpha|}{\sqrt{ (\tilde\Gamma_{P}
   \tilde\Gamma_{S}
   (\tilde\Gamma_{P}
   \tilde\Gamma_{S}+16 G-8)+16)
   \left(\tilde\Gamma_{P}^2+\tilde\Gamma_{S}^2+2
   \sqrt{\tilde\Gamma_{P}
   \tilde\Gamma_{S}
   (\tilde\Gamma_{P}
   \tilde\Gamma_{S}+16
   G-8)+16}-8\right)}}.
   \label{eq:EBP_EA_app}
\ee

\subsection{\hw{Full model of electro-optical transduction}}
\label{app:full_cavity_model_EO}

\color{\revisedfontcolor}

Without loss of generality, here we take the microwave-to-optical transduction as an example, where the signal mode is at microwave band and the probe mode is at optical band. The cavity electro-optical dynamics based on a second-order nonlinear optical medium can be described by the Hamiltonian~\cite{Tsang2010,holzgrafe2020cavity,sahu2022quantum}
\bal  
H&=\hbar \omega_S \hat a_S^\dagger \hat a_S + \hbar \omega_P \hat a_P^\dagger \hat a_P  
% \\&\quad 
- \hbar g \hat a_P^\dagger \hat a_P (\hat a_S+\hat a_S),
\eal
where $\hat a_S,\hat a_P$ are the annihilation operators of the signal (microwave), probe (optical), the frequencies of signal and probe are denoted as $\omega_S, \omega_P$ respectively, the nonlinear coupling coefficient is $g$ in unit of Hz.

Now consider a strong optical pump of mean field $\alpha e^{-i\omega_{\rm pump}t}$ with $\alpha\gg 1$, at frequency $\omega_{\rm pump}=\omega_P+\Delta$ at the red sideband ($\Delta<0$) of the optical probe $\hat a_{P}$, then the interaction Hamiltonian becomes $-\hbar g(\hat{a}_S^{\dagger}+\hat{a}_S) ( \hat{a}_P+\alpha e^{-i\omega_{\rm pump}t})^\dagger(\hat{a}_P+\alpha e^{-i\omega_{\rm pump}t})$. 
Moving $\hat a_P$ into the frame rotating with the optical pump, with the rotating-wave approximation the final Hamiltonian becomes 
\be 
\hat H=\hbar \omega_S \hat a_S^\dagger \hat a_S - \hbar \Delta \hat a_P^\dagger \hat a_P  
-\hbar g(\alpha^* \hat{a}_S^{\dagger}\hat{a}_P+ \alpha\hat{a}_S \hat{a}_P^\dagger).
\ee
In the main text, we considered the ideal case of $\Delta=-\omega_S$, thus one could move $\hat a_S$, $\hat a_P$ into the frame rotating with the cavity resonance frequency instead and obtain the much simpler Hamiltonian \eqref{H_I_EO}.

The input-output relation is described by the Langevin equation~\cite{gardiner1985input,bowen2015quantum}. Below, we summarize the solution of Langevin equation for cavity electro-optical transduction~\cite{holzgrafe2020cavity,sahu2022quantum}.

\subsubsection{Input-output relation}

Consider input field operator vector $\hat \bcalE_{in}\equiv [\hat \calE_{S,in},\hat \calE_{S,E}, \hat \calE_{P,in}, \hat \calE_{P,E}, \hat \calE_{S,in}^\dagger,\hat \calE_{S,E}^\dagger, \hat \calE_{P,in}^\dagger, \hat \calE_{P,E}^\dagger]^T$, where $\hat \calE_{S,in},\hat \calE_{P,in}$ are input fields at signal and probe frequencies respectively, $\hat \calE_{S,E}, \hat \calE_{P,E}$ are environment fields at signal and probe frequencies respectively. And similarly output field operator vector $\hat \bcalE_{out}\equiv [\hat \calE_{S,out}, \hat \calE_{P,out},\hat \calE_{S,out}^\dagger, \hat \calE_{P,out}^\dagger]^T$. Also the cavity mode annihilation operator vector $\hat {\bm a}\equiv [\hat a_S,\hat a_P,\hat a_S^\dagger,\hat a_P^\dagger]^T$.
With the strong optical pump, the Langevin equation is linearized as
\bal
\frac{d}{dt}{\bm{\hat a}}(t) &=  A \hat{\bm a}(t) + B \hat\bcalE_{in} (t)\,, 
\quad 
\hat\bcalE_{out} (t) &= C \hat{\bm a}(t) + D\hat\bcalE_{in} (t)\,,
\eal
where 
\bal 
A&=\left(
\begin{array}{cccc}
 -\frac{\Gamma _S}{2}-i \Delta _S & i g\alpha & 0 & 0 \\
 i g\alpha & -\frac{\Gamma _P}{2}+i \Delta _P & 0 & 0 \\
 0 & 0 & -\frac{\Gamma _S}{2}+i \Delta _S & -i g\alpha^* \\
 0 & 0 & -i g\alpha^* & -\frac{\Gamma _P}{2}-i \Delta _P \\
\end{array}
\right)
\,,
\\
B&=\left(
\begin{array}{cccccccc}
 \sqrt{\gamma _{S,c}} & \sqrt{\gamma _{S,0}} & 0 & 0 & 0 & 0 & 0 & 0 \\
 0 & 0 & \sqrt{\gamma _{P,c}} & \sqrt{\gamma _{P,0}} & 0 & 0 & 0 & 0 \\
 0 & 0 & 0 & 0 & \sqrt{\gamma _{S,c}} & \sqrt{\gamma _{S,0}} & 0 & 0 \\
 0 & 0 & 0 & 0 & 0 & 0 & \sqrt{\gamma _{P,c}} & \sqrt{\gamma _{P,0}} \\
\end{array}
\right)
\,,
\\
C&=\left(
\begin{array}{cccc}
 \sqrt{\gamma _{S,c}} & 0 & 0 & 0 \\
 0 & \sqrt{\gamma _{P,c}} & 0 & 0 \\
 0 & 0 & \sqrt{\gamma _{S,c}} & 0 \\
 0 & 0 & 0 & \sqrt{\gamma _{P,c}} \\
\end{array}
\right)
\,,
\\
D&=\left(
\begin{array}{cccccccc}
 -1 & 0 & 0 & 0 & 0 & 0 & 0 & 0 \\
 0 & 0 & -1 & 0 & 0 & 0 & 0 & 0 \\
 0 & 0 & 0 & 0 & -1 & 0 & 0 & 0 \\
 0 & 0 & 0 & 0 & 0 & 0 & -1 & 0 \\
\end{array}
\right),
\eal
$\gamma_{S,0}, \gamma_{S,c}$ are the intrinsic loss rate and the coupling rate of the signal cavity, similar for $\gamma_{P,0}, \gamma_{P,c}$ of the probe cavity, and $\Gamma_S=\gamma_{S,0}+\gamma_{S,c},\Gamma_P=\gamma_{P,0}+\gamma_{P,c}$ are total linewidths. We define the coupling ratios $\zeta_P\equiv\gamma_{P,c}/\Gamma_{P}, \zeta_S\equiv\gamma_{S,c}/\Gamma_{S}$.
In the steady-state limit, it is convenient to consider the frequency spectrum of the input-output relation.
Fourier transform of the Langevin equation gives
\bal 
\hat\bcalE_{out} (\omega) = S(\omega)\hat\bcalE_{in} (\omega),
\label{eq:input-output_EO_app}
\eal
where $\omega$ is the frequency at the frame rotating with the pump frequency for the optical probe (stationary frame for the microwave signal), the spectral transfer matrix $S(\omega)=C(-i\omega I_4-A)^{-1}B+D$, $I_4$ is a $4\times 4$ identity matrix. Note that here $\omega$ is in a different rotating frame from that in main
text.
From \eqref{eq:input-output_EMO_app}, we obtain the cavity transmissivity for the optical probe 
\be 
\kappa_{\rm oo}(\omega)=|S_{23}(\omega)|^2=
\frac{C^2 \Gamma _P^2 \Gamma _S^2-2 C \Gamma _P \Gamma _S \left(\Gamma _P \left(2 \zeta _P-1\right) \Gamma _S+4 \left(\omega -\omega
   _S\right) \left(\Delta _P+\omega \right)\right)+\left(\Gamma _S^2+4 \left(\omega -\omega _S\right){}^2\right) \left(\Gamma _P^2
   \left(1-2 \zeta _P\right){}^2+4 \left(\Delta _P+\omega \right){}^2\right)}{\Gamma _P^2 \Gamma _S^2 \left| C+\left(1-\frac{2 i
   \left(\omega +\Delta _P\right)}{\Gamma _P}\right) \left(\frac{2 i \left(\omega _S-\omega \right)}{\Gamma _S}+1\right)\right| {}^2},
\ee
and the intrinsic conversion efficiency of electro-optical transduction
\be 
\eta_{\rm eo}(\omega)=|S_{13}(\omega)|^2
=
\frac{4 C \zeta _P \zeta _S}{\left| C+\left(1-\frac{2 i \left(\omega +\Delta _P\right)}{\Gamma _P}\right) \left(\frac{2 i
   \left(\omega _S-\omega \right)}{\Gamma _S}+1\right)\right| {}^2},
\ee
which agree with \eqref{eq:kappa} and \eqref{eq:eta} with $\Delta\to -\omega_S$, $\omega\to \omega+\omega_S$. Below, we evaluate the impact of the pump detuning imperfection $\delta\equiv \Delta+\omega_S$ from the ideal detuning $-\omega_S$.

\subsubsection{Evaluation of the impact of imperfect pump detunings}
\label{app:eval_pumpdetuning}

Fig.~\ref{fig:etaEA_EBPEA_detuned} shows the impact on Fig.~\ref{fig:etaEA_EBPEA} when the pump detuning $\Delta=\omega_{\rm pump}-\omega_P$ deviates from the ideal value $-\omega_S$ by an imperfection $\delta$. We plot $\delta$ ranging from 0 to $40$MHz $(\simeq 3B) $, where $B\approx \Gamma_S=13.706$MHz is the bandwidth of the cavity electro-optical coupler. In subplot (a), we see a decay in the EA efficiency $\eta_{\rm EA}$ with increasing $\delta$, but at a negligible level. Such robustness against detuning imperfection $\delta$ is because the intrinsic loss spectrum $\kappa_E(\omega)=1-\kappa(\omega)-\eta(\omega)$ is centered around the probe cavity resonance frequency $\omega=-\Delta$ ($\omega$ is in the frame rotating with pump) while the electro-optic conversion efficiency spectrum $\eta(\omega)$ is always centered around $\omega=\omega_S$ due to energy conservation, thus the high efficiency region enjoys a smaller $\kappa_E$ for larger detuning imperfection $|\delta|$, and our EA protocol benefits from small $\kappa_E$ as shown in \eqref{eq:etaEA_Ginf}. Such robustness of EA efficiency spectrum $\eta_{\rm EA}(\omega)$ immediately leads to similar robustness of the EA efficiency-bandwidth product and quantum capacity, as shown in subplots (b)(c).

\begin{figure*}[htbp]
    \centering
    \includegraphics[width=0.95\linewidth]{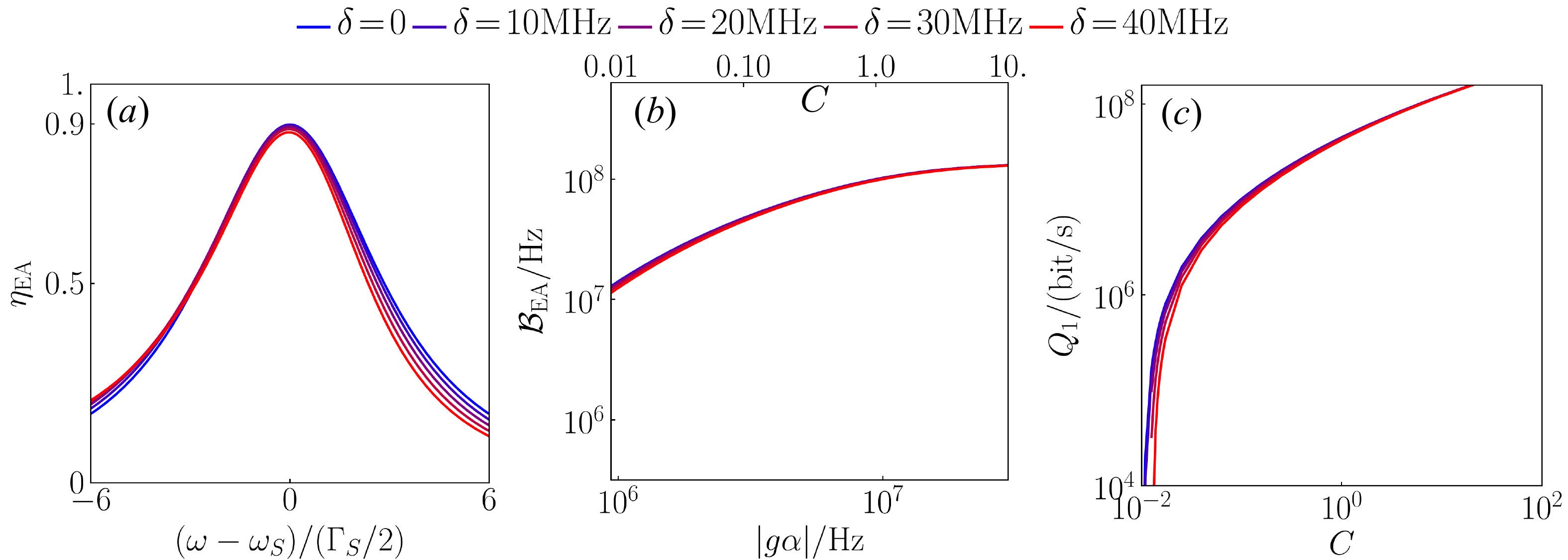}
    \caption{figure}{\hw{Fig.~\ref{fig:etaEA_EBPEA} with pump detuned from the probe cavity resonance frequency by various $\Delta=\omega_{\rm pump}-\omega_P=-\omega_S+\delta$, where $\delta$ is the detuning imperfection from the ideal detuning $-\omega_S$. We consider $\delta$ ranging from 0 to $40$MHz ($\simeq 3\Gamma_S$). Here the squeezer gain is fixed to $G=30$dB. 
    (a) EA transduction efficiency spectrum $\eta_{\rm EA}(\omega)$ under various $\delta$. $C=0.1$. (b) The EA efficiency-bandwidth product $\calB_{\rm EA}$ versus the effective nonlinear coupling strength $|g\alpha|$.
    We fix $\Gamma_P$, $\Gamma_S$, for which we provide the cooperativity $C$ values as the upper axis ticks.
    (c) Broadband quantum capacity rate $Q_1$ versus cooperativity $C$. 
    In all figures, $\zeta_P=\zeta_S=0.99$; Linewidths $\Gamma_P=25.8$MHz, $\Gamma_S=13.706$MHz are chosen according to the high-cooperativity setup in Ref.~\cite{sahu2022quantum}.}
    \label{fig:etaEA_EBPEA_detuned}
    }
\end{figure*}

\subsection{\hw{Implementation of frequency-dependent antisqueezer}}
\label{app:freq_dependent_antisqz}

Here we present a simple design of the frequency-dependent antisqueezer to beat the quantum capacity of the non-EA transduction.

We aim to approach the noiseless antisqueezing gain spectrum $G^{\prime\star}(\omega)$ according to \eqref{eq:optGp_nonideal}. Ideally, $G^{\prime\star}(\omega)\propto 1/\eta(\omega)$ according to \eqref{eq:uniteff} at $G\to\infty$, which is approximately the inverse of the quasi-Lorentzian cavity lineshape $\eta(\omega)$. We propose to adopt a sequential array of cavities to approximate the required spectrum. Here, we consider a specific setup of a 2-periodic array---alternating antisqueezers and squeezers---as shown in Fig.~\ref{fig:freq_dependent_antisqueezer}. We consider a specific class of parametric amplifiers (PAs) of given lineshape as squeezers (and antisqueezers up to $\pi$ phase) with tunable gains and linewidths, and concatenate the squeezers and antisqueezers of various gains and linewidths sequentially together, to approach the desired squeezing lineshape $G^{\prime\star}(\omega)$.

\begin{figure}
    \centering
    \includegraphics[width=0.95\linewidth]{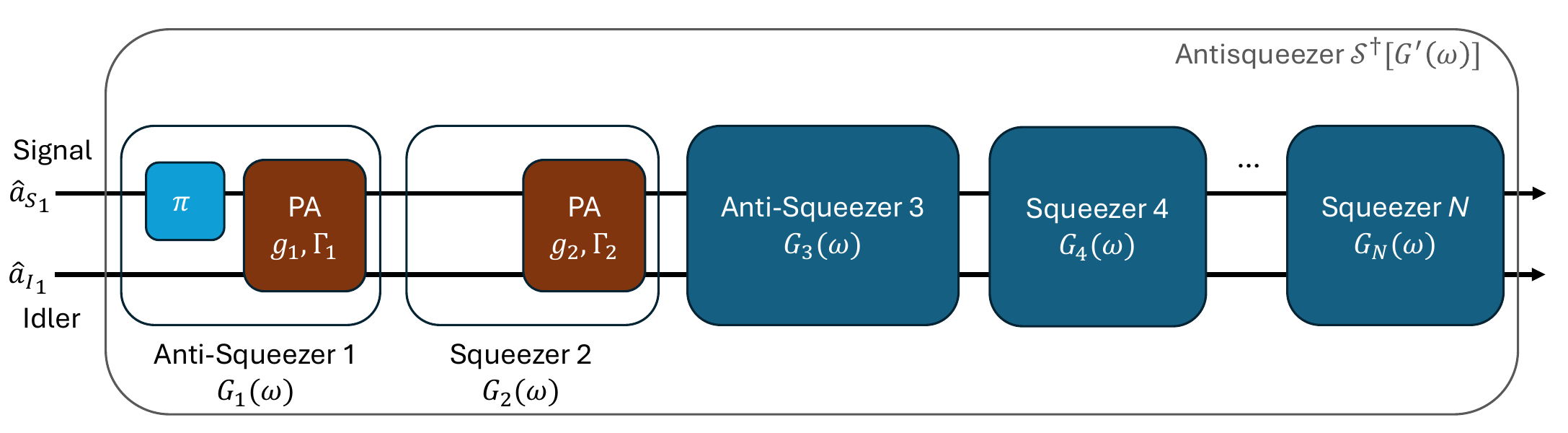}
    \caption{\hw{Schematic of the design of frequency-dependent antisqueezer using $N$-layer $2$-periodic sequential parametric amplifier (PA) array with individually tunable gain spectrum $G_i(\omega)$'s determined by normalized gain $g_i$'s and linewidth $\Gamma_i$'s, $i=1,2,\ldots, N$. Each squeezer component is implemented by a PA, while each antisqueezer component consists of the same type of PA with a $\pi$ phase shift on either signal or idler input port.}
    \label{fig:freq_dependent_antisqueezer}
    }
\end{figure}

\begin{figure}
    \centering
    \includegraphics[width=0.9\linewidth]{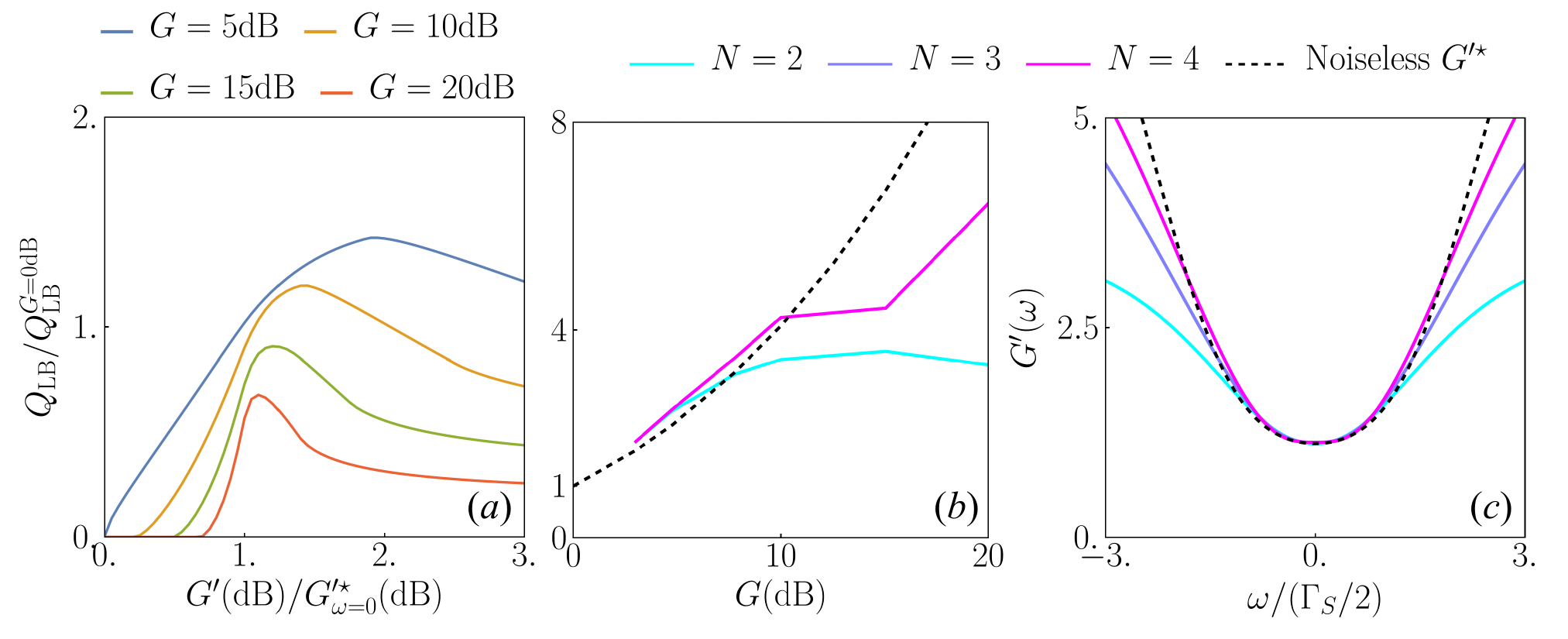}
    \caption{\hw{Performance of various designs of antisqueezer. (a) Control group: EA advantage of an ultra-broadband antisqueezer of constant gain $G'(\omega)=G'$ in $Q_{\rm LB}$ \eqref{eq:Q_LB_broadband_app} over non-EA $Q_{\rm LB}^{G=0{\rm dB}}$, the x axis is normalized by the on-resonance noiseless antisqueezing gain $G^{\prime\star}_{\omega=0}=\frac{1}{1-\kappa(\omega=0)+\kappa(\omega=0)/G}$ (in dB unit); (b) EA advantage of frequency-dependent antisqueezer using $N$-layer $2$-periodic sequential parametric amplifier (PA) array (See Fig.~\ref{fig:freq_dependent_antisqueezer}) with numerically optimized $g_i$'s and $\Gamma_i$'s, compared with the advantage using noiseless antisqueezing $G^{\prime\star}$ (black dashed) which can be surpassed for $G\le 10$dB, since $\eta_{\rm EA}$ can be further increased at the cost of increasing noise; (c) the overall antisqueezing gain spectrum $G'(\omega)$ achieved by the $N$-layer sequential PA arrays, under fixed input squeezing $G=10$dB, compared with the noiseless antisqueezing gain spectrum $G^{\prime\star}(\omega)$ (black dashed). In all subplots, $\kappa(\omega),\eta(\omega)$ are calculated using the parameter setup $\Gamma_P=25.8$MHz, $\Gamma_S=13.706$MHz, chosen from the high-cooperativity setup in Ref.~\cite{sahu2022quantum}. Here we consider ideally overcoupled cavities $\zeta_P=\zeta_S=1$ to accelerate the numerical simulation; and $C=0.49$, higher than $C=0.1$ in the main
    text, since the EA advantage is always infinite as $Q_{\rm LB}^{G=0{\rm dB}}=0$ in the latter case.}
    \label{fig:freq_dependent_antisqueezer_performance}
    }
\end{figure}

We investigate a specific class of doubly-resonant cavity PA~\cite{shapiro2000ultrabright} as an example, which is subject to the Heisenberg-Langevin equations $\frac{d}{dt}\hat a_{S_i}=-\Gamma_i\hat a_{S_i} + g_i\Gamma_i \hat a_{I_i}^\dagger+\sqrt{2\Gamma_i}\hat a_{S_i, in}$, $\frac{d}{dt}\hat a_{I_i}=-\Gamma_i\hat a_{I_i} + g_i\Gamma_i \hat a_{S_i}^\dagger+\sqrt{2\Gamma_i}\hat a_{I_i, in}$ for signal mode $S_i$ and idler mode $I_i$ resonant at the same frequency, where $g_i$, $\Gamma_i$ are the normalized gain (the ratio of pump power over threshold power) and half-linewidth for the $i$th PA. The squeezing gain lineshape is then
\be 
G_i(\omega)=1+\frac{4 g_i^2}{\left(-g_i^2-\frac{\omega^2}{\Gamma_i ^2}+1\right)^2+\frac{4 \omega^2}{\Gamma_i ^2}}\,.
\label{eq:G_OPA_app}
\ee
It is trivial to derive the overall gain for the $N$-layer sequential array iteratively, while the general formula of any $N$ is too lengthy to be presented here. Note that the overall gain is not simply the multiplication of individual gains, due to the interferences with the idler. For $N=2$, the overall gain $G'(\omega)=\left(\sqrt{(G_1(\omega)-1) (G_2(\omega)-1)}-\sqrt{G_1(\omega)G_2(\omega)}\right)^2$. 

An imperfect antisqueezer gain $G'$ invokes additional thermal background of mean photon number
\be 
N_B(\omega)=\left(\sqrt{G \left(G'(\omega)-1\right)}\mp\sqrt{(G-1) \kappa  (\omega)G'(\omega)}\right)^2,
\label{eq:Nb_app}
\ee
where $\mp$ is $-$ for an antisqueezer or $+$ for a squeezer. For $G'(\omega)\to 1,\kappa(\omega)\to 1$, it reduces to the well-known quantum-limited phase-insensitive linear amplification noise $G-1$~\cite{caves1982quantum,Weedbrook2012}. 

A lower bound of the broadband quantum capacity rate of the resulting bosonic thermal loss channel of transmissivity $\eta_{\rm EA}$ and additive thermal background photon number $N_B$ has been presented as \eqref{eq:Q_LB_app} in
Appendix~\ref{app:ancilla_loss}. The broadband rate of it is
\be 
Q_{\rm LB}=\int_{-\infty}^\infty \max\left\{0, \log_2 \left(\frac{\eta_{\rm EA}(\omega)}{\left| 1-\eta_{\rm EA}(\omega)\right| }\right)-g\left(\frac{N_B(\omega)}{\left|
   1-\eta_{\rm EA}(\omega)\right| }\right)
  \right\} \frac{d\omega}{2\pi},
\label{eq:Q_LB_broadband_app}
\ee
which is in unit of bit/s. We use this lower bound to benchmark the performance of our frequency-dependent antisqueezer designs using the sequential PA array.

In Fig.~\ref{fig:freq_dependent_antisqueezer_performance}, we evaluate the performance of the protocol.
 
We begin with a control group, a simple ultra-broadband antisqueezer of uniform gain spectrum, to provide a baseline of the advantage of our proposal of the sequential PA array. Subplot (a) plots the EA advantage in quantum capacity over the non-EA case, with the uniform antisqueezer of constant gain $G'$. We find that the optimal choice of $G'$ close to the noiseless on-resonance value $G^{\prime\star}_{\omega=0}$ as expected. However, the EA advantage is limited to $<2$. In fact, the advantage degrades as the input squeezing $G$ increases from 5dB (blue) to 20dB (red). For $G\ge 15$dB, the EA protocol with such uniform antisqueezer cannot even beat the non-EA case. This is because the quantum amplification noise $N_B(\omega)$ grows with the mismatching between $G'$ and the noiseless gain $G^{\prime\star}(\omega)$ \eqref{eq:optGp_nonideal} more rapidly as $G$ increases, as shown in \eqref{eq:Nb_app}. 
 
Next, we numerically optimize the normalized gain $g_i$'s and linewidth $\Gamma_i$'s of the PAs to maximize the quantum capacity lower bound of \eqref{eq:Q_LB_broadband_app}.
In subplot (b), we observe that the advantage bottleneck due to antisqueezing mismatching is resolved by our sequential PA. Now the EA advantage increases with $G$. With $N=4$, a factor of 6.44 advantage is achieved at $G=20$dB. Further increasing the layer number $N$ may continue to boost the advantage, while the numerical optimization is too costly and we leave it for future study. Remarkably, we observe that the noiseless performance can be surpassed by our sequential PA antisqueezer for $G\le 10$dB, this is not surprising since the EA efficiency $\eta_{\rm EA}(\omega)$ can always be further increased by larger $G'(\omega)$ at the cost of increasing noise $N_B(\omega)$ (which increases with $G$), note that data processing inequality does not apply here because overamplifying (over-antisqueezing) is not a simple Gaussian amplification channel here with the entanglement assistance. In subplot (c), we verify that, with increasing layer number $N$, the optimized overall antisqueezing gain spectrum indeed approaches the inverse-Lorentizian shape $G^{\prime\star}(\omega)\propto 1/\eta(\omega)$ as expected.

\color{black}

\subsection{Electro-optomechanical transduction}
\label{app:full_cavity_model_EMO}

The cavity electro-optomechanical dynamics can be described by the full Hamiltonian~\cite{bowen2015quantum,andrews2014bidirectional}
\bal  
H&=\hbar \omega_S \hat a_S^\dagger \hat a_S + \hbar \omega_P \hat a_P^\dagger \hat a_P + \hbar \omega_M \hat a_M^\dagger \hat a_M 
% \\&\quad 
- \hbar g_S \hat a_S^\dagger \hat a_S \hat x_M- \hbar g_P \hat a_P^\dagger \hat a_P \hat x_M
\eal
where $\hat a_S,\hat a_P,\hat a_M$ are the annihilation operators of the signal (microwave), probe (optical), and mediating (mechanical) modes, $\hat x_M=x_{\rm zp}(\hat a_M +\hat a_M^\dagger)$ with $x_{\rm zp}=\sqrt{\frac{\hbar}{2m\omega_M}}$, the frequencies of signal, probe and mediating modes are denoted as $\omega_S, \omega_P$ and $\omega_M$. Here the nonlinear coupling coefficients $g_S,g_P$ (of electro-mechanical and optomechanical couplings respectively) are in unit of $\rm Hz\cdot m^{-1}$. (In the brackets we take the microwave-to-optical transduction as an example). We define $\calG_{S,P}\equiv g_{S,P}x_{\rm zp}\alpha_{S,P}$ proportional to the nonlinear coupling coefficients and the pumping amplitudes, analogous to $g\alpha$ in the electro-optical coupling~\cite{Tsang2011}. Without loss of generality, we assume $\calG_{S,P}$ real.

The input-output relation is described by the Langevin equation~\cite{gardiner1985input,bowen2015quantum}. Below we summarize the solution of Langevin equation for cavity electro-optomechanical transduction~\cite{andrews2014bidirectional}.

\subsubsection{Input-output relation}

Consider input field operator vector $\hat \bcalE_{in}\equiv [\hat \calE_{S,in},\hat \calE_{S,E}, \hat \calE_{P,in}, \hat \calE_{P,E}, \hat \calE_{M,E},\hat \calE_{S,in}^\dagger,\hat \calE_{S,E}^\dagger, \hat \calE_{P,in}^\dagger, \hat \calE_{P,E}^\dagger, \hat \calE_{M,E}^\dagger]^T$, where $\hat \calE_{S,in},\hat \calE_{P,in}$ are input fields at signal and probe frequencies respectively, $\hat \calE_{S,E}, \hat \calE_{P,E}, \hat \calE_{M,E}$ are environment fields at signal, probe, and mediating frequencies respectively. And similarly output field operator vector $\hat \bcalE_{out}\equiv [\hat \calE_{S,out}, \hat \calE_{P,out},\hat \calE_{S,out}^\dagger, \hat \calE_{P,out}^\dagger]^T$. Also the cavity mode annihilation operator vector $\hat {\bm a}\equiv [\hat a_S,\hat a_P,\hat a_M,\hat a_S^\dagger,\hat a_P^\dagger,\hat a_M^\dagger]^T$.
With strong pumps, the Langevin equation is linearized as
\bal
\frac{d}{dt}{\bm{\hat a}}(t) &=  A \hat{\bm a}(t) + B \hat\bcalE_{in} (t)\,, 
\quad 
\hat\bcalE_{out} (t) &= C \hat{\bm a}(t) + D\hat\bcalE_{in} (t)\,,
\eal
where
\bal 
A&=
\left(
\begin{array}{cccccc}
 -\frac{\Gamma _S}{2}+i \Delta _S & 0 & i {{\cal G}_S} & 0 & 0 & i {{\cal G}_S} \\
 0 & -\frac{\Gamma _P}{2}+i \Delta _P & i {{\cal G}_P} & 0 & 0 & i {{\cal G}_P} \\
 i {{\cal G}_S} & i {{\cal G}_P} & -\frac{\Gamma _M}{2}-i \omega _M & i {{\cal G}_S} & i {{\cal G}_P} & 0 \\
 0 & 0 & -i {{\cal G}_S} & -\frac{\Gamma _S}{2}-i \Delta _S & 0 & -i {{\cal G}_S} \\
 0 & 0 & -i {{\cal G}_P} & 0 & -\frac{\Gamma _P}{2}-i \Delta _P & -i {{\cal G}_P} \\
 -i {{\cal G}_S} & -i {{\cal G}_P} & 0 & -i {{\cal G}_S} & -i {{\cal G}_P} & -\frac{\Gamma _M}{2}+i \omega _M \\
\end{array}
\right)
\,,
\nonumber
\eal
\bal
B&=
\left(
\begin{array}{cccccccccc}
 \sqrt{\gamma _{S,c}} & \sqrt{\gamma _{S,0}} & 0 & 0 & 0 & 0 & 0 & 0 & 0 & 0 \\
 0 & 0 & \sqrt{\gamma _{P,c}} & \sqrt{\gamma _{P,0}} & 0 & 0 & 0 & 0 & 0 & 0 \\
 0 & 0 & 0 & 0 & \sqrt{\Gamma _M} & 0 & 0 & 0 & 0 & 0 \\
 0 & 0 & 0 & 0 & 0 & \sqrt{\gamma _{S,c}} & \sqrt{\gamma _{S,0}} & 0 & 0 & 0 \\
 0 & 0 & 0 & 0 & 0 & 0 & 0 & \sqrt{\gamma _{P,c}} & \sqrt{\gamma _{P,0}} & 0 \\
 0 & 0 & 0 & 0 & 0 & 0 & 0 & 0 & 0 & \sqrt{\Gamma _M} \\
\end{array}
\right)
\,,
\\
C&=
\left(
\begin{array}{cccccc}
 \sqrt{\gamma _{S,c}} & 0 & 0 & 0 & 0 & 0 \\
 0 & \sqrt{\gamma _{P,c}} & 0 & 0 & 0 & 0 \\
 0 & 0 & 0 & \sqrt{\gamma _{S,c}} & 0 & 0 \\
 0 & 0 & 0 & 0 & \sqrt{\gamma _{P,c}} & 0 \\
\end{array}
\right)
\,,\quad 
D=
\left(
\begin{array}{cccccccccc}
 -1 & 0 & 0 & 0 & 0 & 0 & 0 & 0 & 0 & 0 \\
 0 & 0 & -1 & 0 & 0 & 0 & 0 & 0 & 0 & 0 \\
 0 & 0 & 0 & 0 & 0 & -1 & 0 & 0 & 0 & 0 \\
 0 & 0 & 0 & 0 & 0 & 0 & 0 & -1 & 0 & 0 \\
\end{array}
\right)\,.
\eal     
where $\gamma_{S,0}, \gamma_{S,c}$ are the intrinsic loss rate and the coupling rate of the signal cavity, similar for $\gamma_{P,0}, \gamma_{P,c}$ of the probe cavity, total linewidth $\Gamma_S=\gamma_{S,0}+\gamma_{S,c},\Gamma_P=\gamma_{P,0}+\gamma_{P,c}$, $\Delta_S,\Delta_P$ are the detunings of the pumps from the resonance frequencies for signal and probe cavities respectively. We define the coupling ratios $\zeta_P\equiv\gamma_{P,c}/\Gamma_{P}, \zeta_S\equiv\gamma_{S,c}/\Gamma_{S}$.
In the steady-state limit, it is convenient to consider the frequency spectrum of the input-output relation.
Fourier transform of the Langevin equation gives
\bal 
\hat\bcalE_{out} (\omega) = S(\omega)\hat\bcalE_{in} (\omega)
\label{eq:input-output_EMO_app}
\eal
where $\omega$ is the frequency at the frame rotating with the pump frequencies of signal and probe \hw{(stationary frame for the mediating mode)}, the spectral transfer matrix $S(\omega)=C(-i\omega I_6-A)^{-1}B+D$, $I_6$ is a $6\times 6$ identity matrix. \hw{Note that here $\omega$ is in a different rotating frame from the electro-optics model in main text.}

Now consider the red sideband detuning $\Delta_S=\Delta_P=-\omega_M$ to maximize the noiseless beamsplitter-type conversion and suppress the noisy blue sideband squeezing-type conversion~\cite{brubaker2022optomechanical,rau2022entanglement}. From \eqref{eq:input-output_EMO_app}, we obtain the intrinsic conversion efficiency of electro-optomechanical transduction at the frequency resolved limit $\Gamma_S,\Gamma_P\ll\omega_M$:
\bal 
&\eta_{\rm emo}(\omega)\equiv |S_{13}(\omega)|^2 \Big|_{\Delta_S=\Delta_P=-\omega_M; \Gamma_S,\Gamma_P\ll\omega_M}\\
&=\frac{64 {{\cal G}_P}^2 {{\cal G}_S}^2 \Gamma _P \zeta _P \Gamma _S \zeta _S}{4 \Delta\omega ^2 \left(4 {{\cal G}_P}^2+4 {{\cal G}_S}^2+\Gamma _M \Gamma _P+\Gamma _M \Gamma
   _S+\Gamma _P \Gamma _S-4 \Delta\omega ^2\right){}^2+\left(4 {{\cal G}_S}^2 \Gamma _P+4 {{\cal G}_P}^2 \Gamma _S-4 \Delta\omega ^2 \left(\Gamma _M+\Gamma
   _P+\Gamma _S\right)+\Gamma _M \Gamma _P \Gamma _S\right){}^2}
   \label{eq:eta_EMO_app}
\eal  
where $\Delta\omega\equiv \omega-\omega_M$.

\subsubsection{EBP}
\label{app:EBP_EMO}
The transduction efficiency \eqref{eq:eta_EMO_app} has six pure imaginary poles in three pairs $p_1,p_2,p_3, p_4=-p_1,p_5=-p_2,p_6=-p_3$, of which the formulas are too lengthy to be shown here.
Thus, the spectral integral gives the EBP
\bal 
\calB^{\rm emo}&\equiv \int_{-\infty}^\infty \eta_{\rm emo}(\omega) d\omega
=
\frac{i \pi  \left(p_1+p_2+p_3\right) {{\cal G}_P}^2 {{\cal G}_S}^2 \Gamma _P \Gamma _S \zeta _P  \zeta _S}{\left(p_1+p_2\right) \left(p_1+p_3\right)
   \left(p_2+p_3\right) p_1p_2p_3}.
\eal
Here $\calB^{\rm emo}$ is always real as the imaginary sign in the numerator cancels with the imaginary signs of the poles.
We find that $\calB^{\rm emo}$ is maximized at $\Gamma_M\to 0$, given finite cooperativity $C_S=4{{\cal G}_S}^2/\Gamma_S\Gamma_m, C_P=4{{\cal G}_P}^2/\Gamma_P\Gamma_m$. In this case, $p_1\simeq i\frac{\Gamma _M \left(C_P+C_S+1\right)}{2} , p_2\simeq \frac{i \Gamma _P}{2}, p_3\simeq \frac{i \Gamma _S}{2}$, which gives 
\be 
\calB^{\rm emo}|_{\Gamma_M\to 0}\simeq 
\frac{8 \pi  {{\cal G}_P}^2 {{\cal G}_S}^2 \Gamma _P^2 \zeta _P \Gamma _S^2 \zeta _S \left(4 {{\cal G}_P}^2 \Gamma _S+4 {{\cal G}_S}^2 \Gamma _P+\Gamma _P \Gamma _S \left(\Gamma _P+\Gamma _S\right)\right)}{\left(\Gamma _P+\Gamma _S\right) \left({{\cal G}_P}^2 \Gamma _S+{{\cal G}_S}^2 \Gamma _P\right)
   \left(4 {{\cal G}_P}^2 \Gamma _S+\Gamma _P \left(4 {{\cal G}_S}^2+\Gamma _P \Gamma _S\right)\right) \left(4 {{\cal G}_P}^2 \Gamma _S+\Gamma _P \left(4 {{\cal G}_S}^2+\Gamma _S^2\right)\right)}.
\ee

In the symmetric case of ${{\cal G}_P}= {{\cal G}_S}=\calG$, we can obtain the maximum analytically
\be 
\calB^{\rm emo}_{\rm max}= \frac{\sqrt{107+51\sqrt{17}}}{32}\pi \zeta_P\zeta_S {{\cal G}} \simeq 1.749 \zeta_P\zeta_S {{\cal G}}.
\ee 

In the general case of ${{\cal G}_P}\neq {{\cal G}_S}$, exact maximization is in general challenging.
By eliminating terms in the denominator, we find the EBP is upper bounded by
\be 
\calB^{\rm emo}|_{\Gamma_M\to 0}
\le
\pi  \zeta _P \zeta _S \left[4 {{\cal G}_P}^2 \Gamma _S+4 {{\cal G}_S}^2 \Gamma _P+\Gamma _P \Gamma _S \left(\Gamma _P+\Gamma _S\right)\right] \cdot
% \min\left\{\frac{{{\cal G}_P}  }{2 \Gamma _P^2 \Gamma _S},
% \frac{ {{\cal G}_S}  }{2 \Gamma _P \Gamma _S^2}, 
% \frac{1}{32 {{\cal G}_P} {{\cal G}_S}},
% \frac{2 {{\cal G}_P} {{\cal G}_S}}{\Gamma _P^2 \Gamma _S^2}
% \right\}
\min\left\{
\frac{ {{\cal G}_S}^2 \Gamma _P^2}{2 {{\cal G}_P}^4 \Gamma _S^2},
\frac{  {{\cal G}_P}^2 \Gamma _S^2 }{2 {{\cal G}_S}^4 \Gamma _P^2},
   \frac{8  {{\cal G}_S}^2}{\Gamma _S^3 \Gamma _P}
  % \frac{2  {{\cal G}_P}^2 \Gamma _S }{{{\cal G}_S}^2 \Gamma _P^3}
\right\}
\equiv \calB^{\rm emo}_{\rm UB},
\ee
which is maximized to a finite value over any cavity linewidth $\Gamma_S,\Gamma_P$
\be 
\calB^{\rm emo}_{\rm UB,max}=
\frac{4 \pi  \sqrt[8]{{{\cal G}_P}} \zeta _P \zeta _S \left(\left({{\cal G}_P} {{\cal G}_S}\right){}^{3/4}+\sqrt[4]{{{\cal G}_P} {{\cal G}_S}^5}+{{\cal G}_P}^{3/2}+{{\cal G}_S}^{3/2}\right)}{{{\cal G}_S}^{5/8}}
\label{eq:Bmax_EMO_app}
\ee 
at $\Gamma _S\to \frac{2 {{\cal G}_S}^{9/8}}{\sqrt[8]{{{\cal G}_P}}},\Gamma _P\to \frac{2 {{\cal G}_P}^{11/8}}{{{\cal G}_S}^{3/8}}$, where the coupling ratios $\zeta_P,\zeta_S\le 1$. As a reminder, here ${{\cal G}_P},{{\cal G}_S}$ are analogous to $|g\alpha|$ of electro-optical coupling which are independent on cavity quality factor or coupling rate. Hence, similar to the electro-optical transducers, the EBP of electro-optomechanical transducers is fundamentally limited regardless of any cavity engineering.

We note that our result of $\calB^{\rm emo}_{\rm UB,max}$ is not symmetric about $S,P$ since we arbitrarily chose the terms in the denominator of $\calB^{\rm emo}|_{\Gamma_M\to 0}$ to eliminate, which leads to a loose upper bound $\calB^{\rm emo}_{\rm UB}$. More careful choices are likely to offer a tighter upper bound.

\end{appendix}
\end{widetext}
%\bigskip
%\noindent{\bf \normalsize AUTHOR CONTRIBUTION}\\
%\noindent
%HS and QZ proposed the protocol in discussion, performed analyses, generated the figures and wrote the manuscript. 
%
%\bigskip
%\noindent{\bf \normalsize DATA AVAILABILITY}\\
%\noindent
%The data supporting the findings of this study are available from the first author upon reasonable request. 
%
%
%\bigskip
%\noindent{\bf \normalsize CODE AVAILABILITY}\\
%\noindent
%The theoretical results of the manuscript are reproducible from the
%analytical formulas and derivations presented therein. Additional code
%is available from the first author upon reasonable request.
%
%\bigskip
%\noindent{\bf \normalsize COMPETING INTERESTS}\\
%The author declares no competing interests.

%TC:endignore

%\bibliography{ref}
% \bibliographystyle{opticajnl}
% \printbibliography

\end{document}